\documentclass[12pt]{article}
\usepackage[a4paper]{geometry}
\usepackage[english]{babel}
\usepackage[utf8x]{inputenc}
\usepackage{amsmath}
\usepackage{amssymb}
\usepackage{amsthm}
\usepackage{latexsym}
\usepackage{graphicx}
\usepackage{tabularx}
\usepackage{threeparttable} 
\usepackage{caption}
\usepackage{subcaption} 
\usepackage{kpfonts}    
\usepackage{microtype}
\usepackage{booktabs}   
\usepackage{bm}         
\usepackage{listings}   
\usepackage{verbatim}   
\usepackage{color}
\usepackage{stmaryrd}
\usepackage{soul}
\usepackage{hyperref}
\addto\extrasenglish{
    
  }
\hypersetup{            
colorlinks=true
}
\usepackage[colorinlistoftodos]{todonotes}
\usepackage{natbib}
\usepackage{appendix}

\newtheorem{definition}{Definition}[section]

\newcommand{\NN}{\mathbb{N}}
\newcommand{\RR}{\mathbb{R}}

\begin{document}

\title{\textbf{Robust Algorithmic Collusion}\\{\small \textbf{}}}

\author{Nicolas Eschenbaum, Filip Mellgren, Philipp Zahn\thanks{Nicolas Eschenbaum:\ University of St.\thinspace Gallen, Institute of Economics, Varnb\"uelstrasse~19, 9000 St.\thinspace Gallen, Switzerland (nicolas.eschenbaum@unisg.ch); Filip Mellgren:\ Stockholm University, Department of Economics, Universitetsv\"agen 10 A, SE-106 91 Stockholm, Sweden (filip.mellgren@ne.su.se); Philipp Zahn:\ University of St.\thinspace Gallen, Institute of Economics, Varnb\"uelstrasse~19, 9000 St.\thinspace Gallen, Switzerland (philipp.zahn@unisg.ch). Eschenbaum and Zahn gratefully acknowledge financial support from the Basic Research Fund of the University of St. Gallen and the Hasler Foundation.}}
\date{December 2021}
\maketitle

\begin{abstract}
This paper develops a formal framework to assess policies of learning algorithms
in economic games. We investigate whether reinforcement-learning agents with
collusive pricing policies can successfully extrapolate collusive behavior from
training to the market. We find that in testing environments collusion
consistently breaks down. Instead, we observe static Nash play. We then show
that restricting algorithms' strategy space can make algorithmic collusion robust, because it limits overfitting to rival strategies. Our findings suggest that policy-makers should focus on firm behavior aimed at coordinating algorithm design in order to make collusive policies robust.

\end{abstract}

\section{Introduction}
Software systems that take over pricing decisions are becoming widespread. Pricing algorithms can allow firms to monitor and process large amounts of data and adjust prices quickly to changing circumstances. The ascent of such systems poses a potential challenge for the current regulatory landscape: pricing algorithms based on artificial intelligence (AI) may learn to autonomously collude without any previous intentional agreement or explicit instruction to do so.

A growing literature has shown algorithmic collusion to be possible in principle.\footnote{For example, \cite{calvano2020, calvano2021varyingdemand, Klein2021, kastius2021dynamic}.} The results documented so far are a clear warning sign: Even simple algorithms learn to tacitly collude and thereby harm consumers. However, existing analyses have studied the behavior of algorithms in their training environment. It is well-known that algorithms tend to overfit to the training environment, and results cannot easily be extrapolated to other environments \citep[e.g.][]{lanctot2017unified}. In practice, firms train their algorithms offline before using them and face substantial uncertainty about important parameters of the market and their competitors, as well as potentially significant cost from randomized learning in the marketplace. Conclusions drawn from existing work therefore implicitly assume that the training environment and market environment are symmetric and identical, and that results can be extrapolated from one environment to the other.

This paper develops a framework to guide the analysis of learning algorithms in
economic games. We provide a formal representation of the environments that
algorithms face that is parameterized by a ``context''. Evaluating policies in
their training context compared to a (suitably chosen) testing context allows
for an assessment of the behavior of algorithms in markets. We apply our
framework to Q-learning algorithms in repeated Bertrand games. We show that
algorithms overfit to rival strategies from training and cannot successfully
extrapolate their collusive policies to other counterparts, or differently
parameterized environments. In testing contexts algorithmic collusion vanishes,
and does not recover with further iterations. Instead, we observe evidence of
Nash play of the underlying stage game, in particular with (nearly) identically
parameterized environments. Continued policy updating allows algorithms to
overcome this breakdown in collusion, but requires many iterations and is
unlikely to be feasible in market environments. We then show that restricting algorithms' strategy space by only allowing them to condition on their own past price, but not competitors' prices, can make algorithmic collusion robust for a set of parameterizations, because it forces them to learn collusive policies based on simpler patterns that are not too specific to the training context and can thus be successfully extrapolated.

Our findings illustrate a key challenge with the current setup of the analysis of algorithmic collusion: results are reported based on the outcomes in the training environment. In practice however, algorithms are trained and deployed in separate environments. The tendency of machine learning algorithms to overfit to the training environment (or data),  and that therefore a separate testing environment is required to assess their behavior is well-established \citep[see e.g.][]{lanctot2017unified, zhang2018overfitting, zhang2018overfittingcontinuous, song2019overfitting}. But this testing environment is not readily available with reinforcement learners. To the best of our knowledge, we are the first to provide a formal framework to overcome this limitation for algorithms in economic games and develop a consistent approach to assessing behavior.

Our findings highlight the relevance of coordination at the level of algorithm \emph{design}. The tendency of algorithms to jointly learn to collude appears generally insufficient for firms to be able to achieve collusive outcomes in the market. However, they may still be able to successfully achieve algorithmic (tacit) collusion by coordinating on high-level approaches to the implementation of pricing algorithms. Each firm implements and trains its algorithm independently, yet by having coordinated parts of the parameterization and the strategy space appropriately, collusive policies robust to deployment in the market can be consistently learnt. Intuitively, because the extent of coordination among competing firms is (legally) restricted, and algorithms may need to work in a range of environments, the policies employed must rely on simpler patterns.

This paper contributes to a growing literature on algorithmic collusion (for a recent survey of the economic literature on AI see \citet{Abrardi2021survey}). We employ simple Q-learning algorithms in line with related work in e.g. \citet{calvano2020, calvano2021varyingdemand, Klein2021}. Our baseline scenario and parameterization is built on the environment studied in \cite{calvano2020}. There is also a related and growing literature on reinforcement learning in revenue management practice, demonstrating the relevance of assessing reinforcement learners in economic games \citep[e.g.][]{kastius2021dynamic, acuna2021price, bondoux2020reinforcement}.

Our paper is related to the computer science literature that studies the overfitting of reinforcement learning (RL) algorithms. \citet{lanctot2017unified} show that the overfitting to rival agents' policies we observe is a common problem in RL. \citet{zhang2018overfitting} examine different ways how deep RL algorithms overfit to the environment and show that attempted solutions in the literature of adding stochasticity to the environment do not necessarily prevent overfitting. The difficulty for algorithms to extrapolate policies to new environments is known as the ``zero-shot coordination problem'' \citep[e.g.][]{treutlein2021zeroshotcoordination, hu2020otherplay}. Our framework builds on \cite{kirkSurveyGeneralisationDeep2021}. Related to our approach is a strand of literature that assumes there exists a distribution of Markov-decision-problems of the scenario of interest, and then trains algorithms on a finite set of samples from this distribution before testing the behavior on the entire distribution \citep[e.g.][]{zhang2018overfittingcontinuous, nichol2018gotta, justesen2018illuminating}.

Lastly, our paper contributes to the literature on competition policy and regulatory responses to algorithmic collusion. The potential challenge to policy has been previously discussed both by the European Commissioner for Competition \citep{vestager2017algosberlin} and Commissioner of the Federal Trade Commission \citep{ohlhausen2017lawalgos}, and potential solutions have been suggested \citep[e.g.][]{Calvano2020Science, harrington2018developing, beneke2021remedies}. Our results provide some novel perspectives and qualify existing results. When learning in the market itself is feasible and not too costly, algorithmic collusion appears very likely to arise. In any other case, our findings suggest that the main policy challenge is to detect or prevent coordination of algorithm design, implying that the actual danger of algorithmic collusion may not necessarily be in the market interaction itself, but in coordinative moves beforehand.\footnote{One particular case of this, where competing firms buy pricing-services from the same upstream supplier, has been noted before in the literature. For instance, \citet{harrington2021effect} develops a model of sellers that outsource their pricing algorithms to a third-party.} For instance, pricing specialists at rival firms may be well-informed about the work of their counterparts and there may be an industry-level understanding of the best specification of pricing algorithms.

The remainder of this paper is organized as follows. In \autoref{sec:training_vs_testing_model} we develop a formal framework for the analysis of learners in testing environments. \autoref{sec:model} introduces the specific setting of algorithmic collusion that we study, and in \autoref{subsec:test-training} details how we apply our framework in this setting. \autoref{sec:results} presents and discusses our results. Finally, \autoref{sec:conclusion} concludes.

\section{Training vs. Testing: A General Framework}\label{sec:training_vs_testing_model}

The evidence accumulated on the behavior of reinforcement learners in economic games is based on scenarios where the agents learn together in the same environment. In the simplest scenario, two learners are interacting iteratively until either convergence or a fixed number of iterations is reached. The assessment of pricing behavior is then based on the last rounds of these interactions. In practice however, firms train their algorithms offline first before deploying them to the market. Algorithms must therefore successfully extrapolate collusive policies from training to the market, and evidence of collusion during training does not imply that firms can actually employ learning systems that tacitly collude. Instead, an assessment of the behavior of algorithms requires a separation between ``training'' and  ``testing'' environments.\footnote{This is a general challenge in the reinforcement learning literature, and is also known as the ``zero-shot coordination problem'' \citep[see e.g.][]{treutlein2021zeroshotcoordination, hu2020otherplay}.}

In this section, we develop a general framework to guide the analysis of learners in testing environments. We provide a formal representation of the interactive environments that learning agents face. We consider these environments to be parameterized by some ``context''. This context describes how the environment varies with a change in exogenous conditions. Evaluating the difference in behavior between the training and testing environment is then equivalent to comparing the performance of an algorithm between two contexts, one for training and one for testing.

We begin by defining the dynamic system in which interactive reinforcement
learning takes place. Scenarios of interest for economists can be cast in terms
of a \emph{Partially Observable Markov Game}.\footnote{In the economic
  literature, these games are typically referred to as Stochastic Games. Here,
  we follow the terminology that is standard in the machine learning
  literature.} Formally:

\begin{definition}[Partially Observable Markov Game (POMG)]\label{def-pomg}
A Partially Observable Markov Game is formalized by the tuple $(\mathcal{N},S,A,O,T,R,s_0,o_0,\delta)$. Where:\\
$\mathcal{N} = \{ 1,...,N\}$ is the set of agents.\\
$S$ is the set of unobservable states, and $s_0\in S$ the initial state.\\
$A$ is the joint action space $A=A_1 \times ... \times A_N$.\\
$O$ is the joint set of observations $O=O_1 \times ... \times O_N$, and $o_0 \in O$ the initial observation.\\
$\tau$ is the transition probability $\tau \colon S \times A \to \Delta(S,O)$.\\
$R_i$ is the reward function of a player $i \in 1,...,N$, $R_i \colon S \times A \times S \to \RR $.\\
$\delta$ is the common discount factor.
\end{definition}

In a \emph{partially observable} game, the players may not be able to observe the underlying state of the game. We model this by introducing a joint set of observations $O$ and observation profiles $o=(o_1,...,o_N) \in O$, with the initial observation being $o_0 \in O$. Each period $t \in 0,...,T$, the players choose actions $a=(a_1,...,a_N)$ and subsequently transition to the next state. The probability that joint action $a$ in state $s$ leads to a transition to state $s'$ and observation $o'$ is $\tau (s',o'|s,a)$.

The goal in reinforcement learning is for each agent $i$ to  learn a policy
$b_i(a_i|o_i)$ which produces a
probability distribution over actions given an observation, such that the cumulative, discounted reward of the policy is maximized.

In order to distinguish different environments, we follow the ideas in \cite{kirkSurveyGeneralisationDeep2021} and consider a \emph{Contextual Partially Observable Markov Game}, where we introduce a set of contexts $K$.\footnote{Note that \cite{kirkSurveyGeneralisationDeep2021} focus on a single agent and not interactions of players. Their formalization is therefore based on Partially Observable Markov Decision Problems.} For each context $k \in K$ we have a Partially Observable Markov Game (POMG) with the property that the state of the game can be decomposed into two parts, $s = (k, s') \in S^k$, where $s' \in S$ is the state and $k \in K$ is the context. Formally:

\begin{definition}[Contextual Partially Observable Markov Game (CPOMG)]\label{def-cpomg}
  A Contextual Partially Observable Markov Game is formalized by the tuple
  $(N,S,A,O,\allowbreak \tau,R,s_0,o_0,\delta,K)$ where $K$ is a set of
  contexts. $K$ introduces a collection of POMGs. That is, for each $k\in K$ we have a POMG with $(N,S^k,A,O,\tau,R,s^k_0,o^k_0,\delta)$ with the property that the state of that game can be decomposed into two parts: $s = (k,s') \in S^k$. $s'\in S$ is the state and $k \in K$ is the context.
\end{definition}

In contrast to the state $s$, which evolves over the course of the game, we assume that $k$ is fixed throughout. Thus, each $k \in K$ yields a different POMG. The context can capture a variety of aspects. In our application, we will consider two types of different contexts: the initial seeds of a game, and the parameters of the reward function.

As each context induces a separate POMG, we consider separate learning scenarios for each one of these. For each scenario in turn, a learning algorithm will return a policy for each player. We therefore index the policy for a player $i$ by its context $k$, i.e. $b_i^k(a_i|o_i)$. We can then use these learned policies to evaluate their performance in a specific context $k$ of the CPOMG $G$, which may be different from the context in which the policy was learned. We denote the performance for player $i$ as $\mathcal{M}_{i}(b_i^{k_i}(\cdot), b_{-i}^{k_{-i}}(\cdot),G_{k})$, where $b_{-i}^{k_{-i}}(\cdot)$ denotes the policies of all players other than player $i$. The evaluation for player $i$ hinges on the policy he learned in context $k_i$, the policies of his opponents learned in their contexts $k_{-i}$ and on the context of the environment $k$.

Consider the following example (a version of which we analyze later in detail) with two contexts, $k'$ and $k''$. After letting two algorithms jointly learn in each specific context, we observe the rewards of the induced policies for a fixed number of periods. Denote these policies by $b_1^{k^{'}}(\cdot)$ and $b_2^{k^{'}}(\cdot)$ for a context $k'$, and by $b_1^{k^{''}}(\cdot)$ and $b_2^{k^{''}}(\cdot)$ for a context $k''$. We can then evaluate the rewards for player $1$ in the context in which he learned the policy, $\mathcal{M}_{1}(b_1^{k'}(\cdot), b_{2}^{k'}(\cdot),G_{k'})$, and equally in the context  in which he did not learn,  $\mathcal{M}_{1}(b_1^{k'}(\cdot), b_{2}^{k''}(\cdot),G_{k''})$.

Thus, by comparing the performance in different contexts, we can assess the extent to which algorithms are able to extrapolate policies learned in one context to a different context. Selecting contexts appropriately in order to match the decision-making problem firms face in practice then allows us to obtain an assessment of the ability of algorithms to lead to algorithmic collusion in the market. In \autoref{subsec:test-training} we give details on the specific comparisons we consider in this paper.

We conclude this section with a remark on the definitions above. In this paper,
we are comparing ``fixed'' policies across given contexts. But in principle the
framework we introduce can be extended so that the learning algorithms
themselves could condition on the different scenarios. To see this, note that if
we added a probability distribution on the contexts $k$, the CPOMG itself
becomes a well defined POMG. The learners then engage in a ``meta-learning''
problem with awareness of the contexts.  This is the background of the framework
introduced in \cite{kirkSurveyGeneralisationDeep2021}. Similarly, we will
consider a discrete set of contexts in our analysis here. Instead, one could
consider a distribution of contexts. It is straightforward to extend our
framework in this direction.

\section{Model Specification}\label{sec:model}

We now introduce the specific setting we study in our application to algorithmic collusion.

\subsection{Players, Reward Function, and Economic Model}\label{subsec:econmodel}

We model the game-theoretic environment as a standard repeated Bertrand setting. The specification of the demand function and baseline parameterizations are in line with the setup employed in \citet{calvano2020}. This provides a benchmark of existing findings of algorithmic collusion that our results can be directly compared to.

We consider a Bertrand game with two players in our baseline setup, $N=2$. Players choose prices as actions in each period $t \in 0,...,T$. We let $p_{i,t}, p_{j,t}$ denote the period-$t$ prices of players $i$ and $j$ respectively, where $i \neq j$, $i,j \in \{1,2\}$. The corresponding quantities are denoted by $q_{i,t}, q_{j,t}$, and marginal cost of players by $c_i, c_j$. The demand function is given by a classic logit-demand specification of
$$q_{i,t} = \frac{e^{\frac{\gamma_i-p_{i,t}}{\mu}}}{\sum_{j=1}^n e^{\frac{\gamma_j - p_{j,t}}{\mu}} + e^{\frac{\gamma_0}{\mu}} } \text{,}$$
where $\gamma_{i}$ denotes the quality parameter of the good supplied by firm $i$ (vertical differentiation) with $\gamma_0 = 0$ being the product quality of the outside good, and $\mu$ the index of horizontal differentiation, so that the goods are perfect substitutes in the limit when $\mu \rightarrow \infty$. For our baseline parameterization we set $\gamma_i - c_i = 1$ and $\mu = 1/4$

The per-period reward of each player is equal to the per-period profit obtained,
$$
R_{i,t} = \pi_{i,t} =  (p_{i,t} - c_i)q_{i,t} \text{.}
$$

\subsection{Action Space}\label{subsec:actionspace}

The learning model we consider requires a discretization of the action space. We therefore consider a discretized joint action space $A$. We construct the grid of prices each player $i$ can choose from, $A_i$, as follows.

Let the set of one-shot Nash equilibrium prices corresponding to all
parameterizations we consider be $\mathbf{p}^N$ and similarly let the set of
joint profit maximizing prices be $\mathbf{p}^C$. Then the grid of prices
available, $A_i$, is given by $k \in \NN $ equally spaced points in the interval
$[\mathbf{\underline{p}}^N - \xi(\mathbf{\bar{p}}^C-\mathbf{\underline{p}}^N),
\mathbf{\bar{p}}^C + \xi(\mathbf{\bar{p}}^C-\mathbf{\underline{p}}^N)]$ with
$\xi > 0$, where $\mathbf{\underline{p}}^N = [\min\{p \in \mathbf{p}^N\}]^2$ and
$\mathbf{\bar{p}}^C = [\max\{p \in \mathbf{p}^C\}]^2$. For our main
specifications we set $k = 20$ and $\xi = 0.1$. Note, the choice of demand
parameters discussed in \autoref{subsec:econmodel} and \autoref{subsec:test-training} imply $\underline{\mathbf{p}}^N
\approx (1.47, 1.47)$ and $\bar{\mathbf{p}}^C \approx (2.62, 2.62)$.

\subsection{States, Transitions, and Observations}

The set of states $S$ of the game is given by the set of price profiles
$(p_{1,t}, p_{2,t})$ and the context, $S = A_1 \times A_2 \times K = A \times
K$. In this paper, we consider a deterministic transition function. As $k\in K$
is fixed, at any time $t$, the next state $s_{t+1}$ is deterministically given
by the actions of players at time $t$, $(p_{1,t}, p_{2,t},k)$.

For our baseline parametization, players have the same set of observations
$O_i$. While they cannot observe $k$, they observe the last period's prices. $O
= O_1 \times O_2 = (A_1 \times A_2) \times (A_1 \times A_2)$. The observation
profile at any time $t$ is therefore $o_t=(o_{1,t} = (p_{1,t-1}, p_{2,t-1}),
o_{2,t} = (p_{1,t-1}, p_{2,t-1}))$. In other words, players have a one-period long memory.

\subsection{Learning Model}\label{subsec:learningmodel}

We employ Q-learning as our learning model, a standard model-free reinforcement learning algorithm.
The algorithm computes a so-called Q-function of expected rewards for an action taken for a given observation -- $Q : O \times A \rightarrow \mathbb{R}$. Q-learning stores the (current) computed Q-value of each observation-action pair in a table and hence requires the action and observation space to be discrete. In each period, the  cell in this Q-matrix corresponding to the current period's observation-action combination, $Q_t(o_{i,t},a_{i,t})$, is updated based on the observed reward in the current period, $R_{i,t}(a_{i,t},a_{j,t}))$, and a learning rate $\alpha \in [0,1]$, according to the following Bellman equation
$$Q_{i,t+1} (o_{i,t},a_{i,t}) = (1-\alpha)Q_{i,t}(o_{i,t},a_{i,t}) + \alpha (R_{i,t} + \delta \max_{a \in A} Q_{i,t}(o_{i,t+1},a)) \text{,}$$
where $\delta$ is the discount factor.

The initial state and the initial Q-matrix must be specified by the programmer
at the start of the learning process. We initialize the Q-matrix with the
Q-values that would arise if both agents randomized uniformly. We also choose the initial state at random.

In each period, the algorithm chooses an action (a price) either in order to \emph{explore} the environment or to \emph{exploit} its current state of knowledge. When exploring, the agent chooses an action at random. When exploiting, it chooses the action with the highest Q-value in the current state. We employ standard $\varepsilon_t$- greedy exploration in which the agent explores with probability $\varepsilon_t$ and exploits with probability $1-\varepsilon_t$. We let $\varepsilon_t$ vary according to $\varepsilon_t = e^{-\beta t}$ where $\beta = 4 \times 10^{-6}$, implying that agents explore relatively often at the start and focus on exploiting over time. We focus on a learning rate  of $\alpha = 0.15$.

We stop the learning process when we observe that the algorithms have converged. Specifically, a given run is stopped if for each player $i$ the action $a_{i,t} = \arg \max\{Q_{i, t}(a, o_t)\}$ does not change for $100 000$ consecutive periods, or after one billion total repetitions. We obtain convergence for over 99 percent of runs.

\subsection{Outcome Measures}\label{subsec:outcomes}

To assess the propensity of algorithms to collude, we focus on two measures of rewards (or profits): the collusion index $M$ and the profit gain $\Delta$. Both express the realized reward in relation to the static Nash equilibrium and joint profit maximizing profits. Specifically, the two metrics are defined as
$$ M = \frac{\bar{\pi} - \pi^N}{\pi^C - \pi^N} \text{,}$$
$$ \Delta_i = \frac{\bar{\pi}_i - \pi_i^N}{\pi_i^C - \pi_i^N} \text{,}$$
where $\bar{\pi}_i$ denotes the average reward (profit) of agent $i$, $\pi_i^N$ the profit of agent $i$ in the one-shot Nash equilibrium of the game, and $\pi_i^C$ the profit of agent $i$ in the joint profit maximizing outcome of the one-shot game. $\bar{\pi}$, $\pi^N$, and $\pi^C$ are defined analogously, but always represent the (average of the) \emph{sum} of profits of players $i$ and $j \neq i$, $i,j \in \{1,2\}$.

Thus, for both the collusion index and an individual player's profit gain, when the respective measure is zero the average profit is equal to the Nash profit, while a value of one implies that the average profit is equal to the joint profit maximizing profits. Note that by definition, the collusion index is equal to the average of the profit gains of the two players.

In addition, we investigate the actions played by agents in more detail. We
classify outcomes based on the unique convergence to specific actions. If both
agents choose the same price in more than 90\% of rounds, we classify the
outcome as \emph{symmetric} convergence. If both agents choose a unique but different price in more than 90\% of rounds, we classify the outcome as \emph{asymmetric} convergence. Finally, if at least one agent plays the same sequence of prices repeatedly in over 90\% of rounds, we classify the outcome as a price \emph{cycle}.\footnote{In principle, we can search for price cycles of any length. However, in practice we limit ourselves to a search for cycles with a maximum length of 15 to avoid costly computations.} If we cannot identify either unique convergence or a price cycle, we classify the outcome as \emph{other}. The vast majority of our learning runs converge to unique symmetric or asymmetric prices and we barely observe any longer price cycles.

\subsection{The Context, and Training vs. Testing Environments}\label{subsec:test-training}

We study a set of fixed contexts for the setting described above. We consider the context to define two aspects of the resulting POMG: the seed used to initialize the learning process, and the parameters of the reward function. We focus on these two aspects to capture the underlying decision-making problem of the firm. When training offline before deploying the algorithm to the market, it is necessarily the case that the rival player changes from training to testing
environment. Thus, by comparing the policies learned in two different POMGs, in which only initial seeds differed, we can analyze the behavior of an algorithm when faced with a previously unseen competitor in the market, but all else remains constant. Similarly, firms face uncertainty over important parameters of rival firms or the demand function, and by comparing policies from two POMGs in which parameters of the reward function are different, we can similarly observe whether players are able to extrapolate potentially collusive policies to the market environment, which might differ in some aspects to the training environment.

We proceed as described in the example before, and always contrast two contexts. For example, let $k'$ and $k''$ be two contexts that yield the same POMG, except for the initial seed. We initialize both learning processes and let two pairs of algorithms learn in their respective context. Denote the policies these algorithms converge to by $b_1^{k^{'}}(\cdot)$ and $b_2^{k^{'}}(\cdot)$ for context $k'$, and by $b_1^{k^{''}}(\cdot)$ and $b_2^{k^{''}}(\cdot)$ for context $k''$. To evaluate the policies, we consider a third context $k'''$ with the same paramterization as contexts $k'$ and $k''$, and a random initial state and new random seed. We then evaluate the outcome measures described in \autoref{subsec:outcomes} for the policy of each player $i \in \{1,2\}$ in their respective training contexts, $\mathcal{M}_{i}(b_i^{k'}(\cdot), b_{-i}^{k'}(\cdot),G_{k'})$, and when two learnt policies from two different training contexts are placed in the new, third context, $\mathcal{M}_{i}(b_i^{k'}(\cdot), b_{-i}^{k''}(\cdot),G_{k'''})$.\footnote{Note that in general we must introduce a third context, as the two initial contexts may have different seeds (requiring us to pick a seed when matching the two resulting policies), and may have converged to different states (requiring us to choose an initial state). However, we will study an environment in \autoref{subsec:robust} in which no new initial state is required.}
We further study the policies learnt in more detail to understand the underlying strategies that yield the outcomes we observe.

The different parameterizations of the reward function that the set of contexts we consider defines vary in the marginal cost level $c$. This mimicks the uncertainty firms may face regarding rival parameters, even when they are perfectly informed about the demand environment. We consider variation in constant marginal cost levels in the range $c_i, c_j \in [1, 1.7]$. We stop at $1.7$ to ensure that either players' cost remain below the monopoly price of a seller with the lowest cost $c=1$, so that it is never optimal for the market to be served by only one firm.
Our specification of demand is particularly well-suited for contexts that yield variation across cost levels. It implies that both the Nash equilibrium profits and the joint profit maximizing profits are constant across all cost levels and all combinations of cost, since $\gamma_i - c_i = 1$. Only the associated optimal prices change. Thus, when algorithms are placed in a new context with a different-cost competitor than previously, no change in the equilibrium price is required by the agent. Only a strategy to \emph{support} high, supra-competitive profits will require a different set of prices, but can be obtained by appropriately `shifting' the prices along the grid in line with the change in cost of the competitor. Hence, the challenge the algorithm faces in the new context is particularly simple and the profits that can be obtained remain constant, ensuring that differences in the outcome observed are not due to a change in possible payoffs for the players.

In the new context (i.e. testing environments), no learning takes place initially and algorithms are only exploiting, i.e. choosing the optimal action given their Q-matrix. They are thus employing their ``fixed'' strategy learnt during training. We further investigate the outcome when agents may continue to learn `in the market', and thus can converge to a new policy.
We ensure that the number of sessions in testing contexts is equal to the number of sessions in training contexts by always pairing policies from the same respective session number.\footnote{Note that the sessions per context run independently of one another. The session number has no further implications, but since there are an exponential number of possible player-context-session-number matches, using it is a straightforward way to limit computations.}

\section{Results}\label{sec:results}

\subsection{Algorithms in Training}\label{subsec:learning}

We begin with an examination of learnt policies in their training context. The contexts vary in two dimensions, as detailed before. We aggregate our results across the variation in initial seeds for each parameterization of the environment. Thus, this is essentially an extension of previous work by \cite{calvano2020} across a range of parameterizations.

We provide summary statistics for the training contexts in the Appendix in \autoref{tab:sumstatsphase1}. Across all parameterizations, we observe a high collusion index between 0.70 and 0.87 and predominantly convergence to unique prices for each player. Profit gains are similarly high and symmetric. The average collusion index for the low and intermediate cost contexts in particular is in line with previous estimates, while for high cost contexts we observe slightly lower values.

The definition of the price grid that we consider implies that the high profits we observe for low cost contexts is somewhat unsurprising. For low cost contexts almost all available prices lie above the static Nash equilibrium price and on average these yield a higher profit than the Nash equilibrium. For high cost contexts this is not the case, since the Nash price already lies in the upper part of the price grid and thus most prices in the grid yield below-Nash profits. For intermediate and high cost contexts our setup thus captures the trade-off that firms face: employing pricing algorithms may be costly in the short run due to exploration, but may pay off in the long run due to above-Nash (and potentially collusive) play when exploiting. These short-term costs can be avoided by training offline first. For low cost contexts instead, costs of exploration are much less of an issue and we would therefore expect that in these contexts agents may learn to converge to above-Nash prices when employing a reinforcement learning algorithm.\footnote{The importance of exploration cost in practice have been noted before by revenue management practitioners, e.g. \cite{kastius2021dynamic}.}

To illustrate this, \autoref{tab:costoflearning} in the Appendix shows the profit loss relative to Nash play  across the grid of cost levels when both players randomize. Players in contexts with a high cost level consistently lose from randomizing, compared to playing the unique Nash equilibrium. But they also lose if the opponent randomizes and the agent itself plays Nash or the best-response to random play. Players in low cost contexts on the other hand \emph{benefit} from randomized play and achieve above-Nash profits. The results we observe for high cost contexts thus show that findings of algorithmic collusion extend to and are stable in environments in which exploration is costly, and short-run costs from learning must be balanced by long-term benefits from algorithmic collusion.

\begin{figure}[h]
  \centering
  \caption{Average path of play following deviation to Nash}
  \includegraphics[width=0.9\columnwidth]{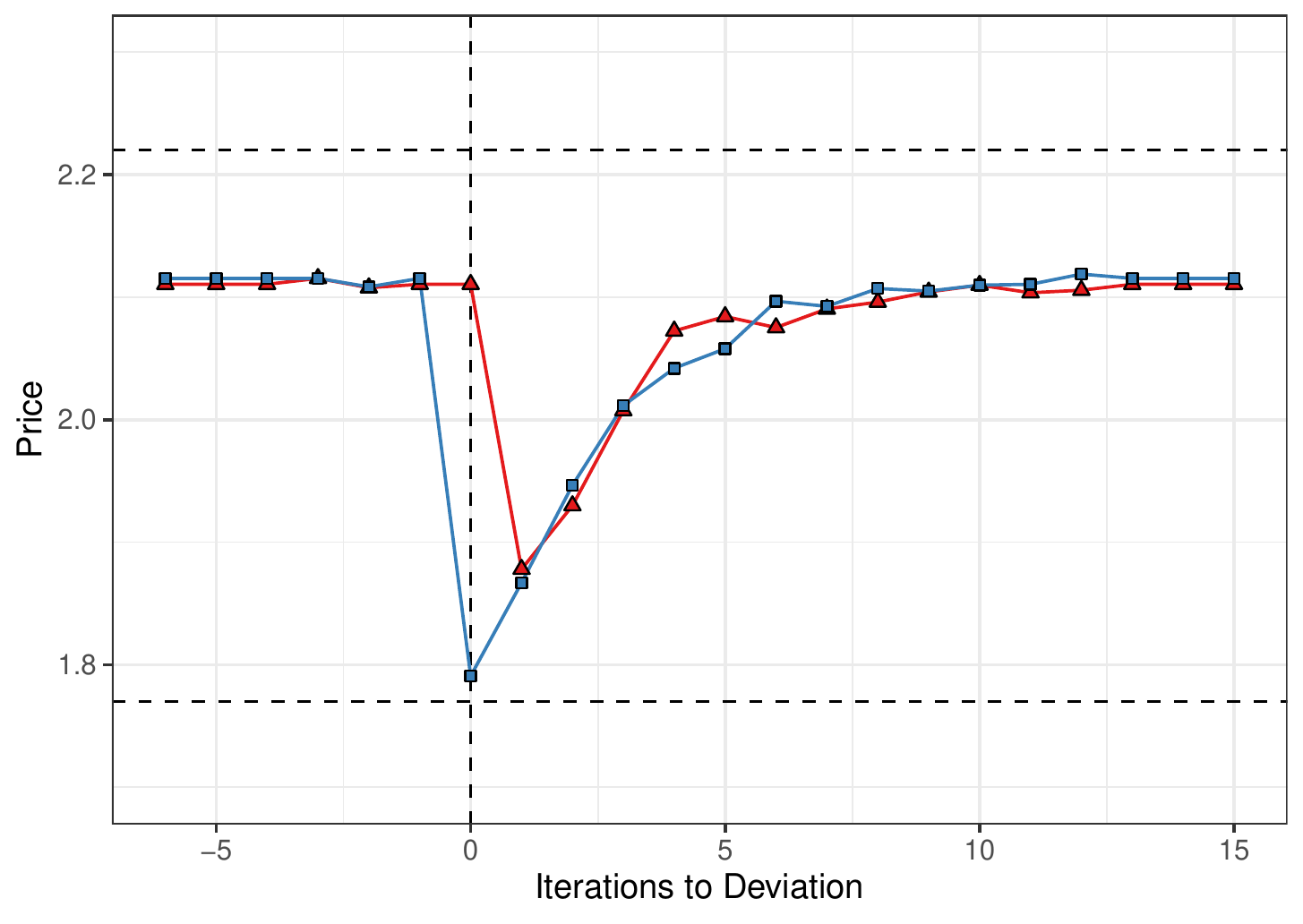}\\ \vspace{-1.5em}
  \flushleft{\footnotesize\textbf{Notes:} The figure shows the average actions
    played prior to and following a manual deviation of one player
    ({\color{blue}{in blue}}) to the price closest to the Nash price on the grid for training contexts with intermediate parameterization. The horizontal dashed lines indicate the static Nash and joint profit maximizing prices.}
  \label{fig:deviationpunishment}
\end{figure}

To confirm that these high, seemingly-collusive outcomes we observe are indeed evidence of algorithmic collusion, we investigate the policies algorithms converged to. \autoref{fig:deviationpunishment} shows for the case of an intermediate cost context the average prices played following a forced deviation to the price nearest to the Nash price on the grid. In line with the literature, we observe a path of play in which the deviating algorithm is punished for the deviation and a gradual return to the stable pre-deviation prices. We observe very similar patterns across the different training contexts.

Lastly, \autoref{fig:convergencetime} in the Appendix shows the time to convergence. Algorithms consistently require more than 1 million rounds of play and on average over 2 million rounds
to achieve convergence. Thus, learning online is likely infeasible in practice. In light of the potential high per-period cost for the firm from exploration, there may be hundreds of thousands of periods of significant losses before the algorithm begins achieving supra-Nash profits. Thus, a separation between training and testing is economically relevant.

\subsection{Algorithms in New Contexts}\label{subsec:testing}

We now assess the performance of learnt policies in contexts that are different
to the training context. We begin by assessing policies that were learnt in two
contexts which only differed in their initial seed in a new context with the
same parameterization.

Recall (and as explained in more detail in \autoref{subsec:test-training}): We
consider converged policies from two contexts, $k'$ and $k''$, denoted by $b_1^{k^{'}}(\cdot)$, $b_2^{k^{'}}(\cdot)$, and $b_1^{k^{''}}(\cdot)$,  $b_2^{k^{''}}(\cdot)$ respectively. To evaluate the policies, we consider a third context $k'''$ with the same parameterization as contexts $k'$ and $k''$, and a random initial state and new random seed. We then evaluate when two learnt policies from two different training contexts are placed in the new, third context, $\mathcal{M}_{i}(b_i^{k'}(\cdot), b_{-i}^{k''}(\cdot),G_{k'''})$. The sole difference for algorithms is therefore that they are facing a so far unseen competitor and employ their ``fixed'', previously learnt policy.

\begin{figure}[h]
  \centering
  \caption{Collusion index in training and testing contexts}
  \includegraphics[width=0.9\columnwidth]{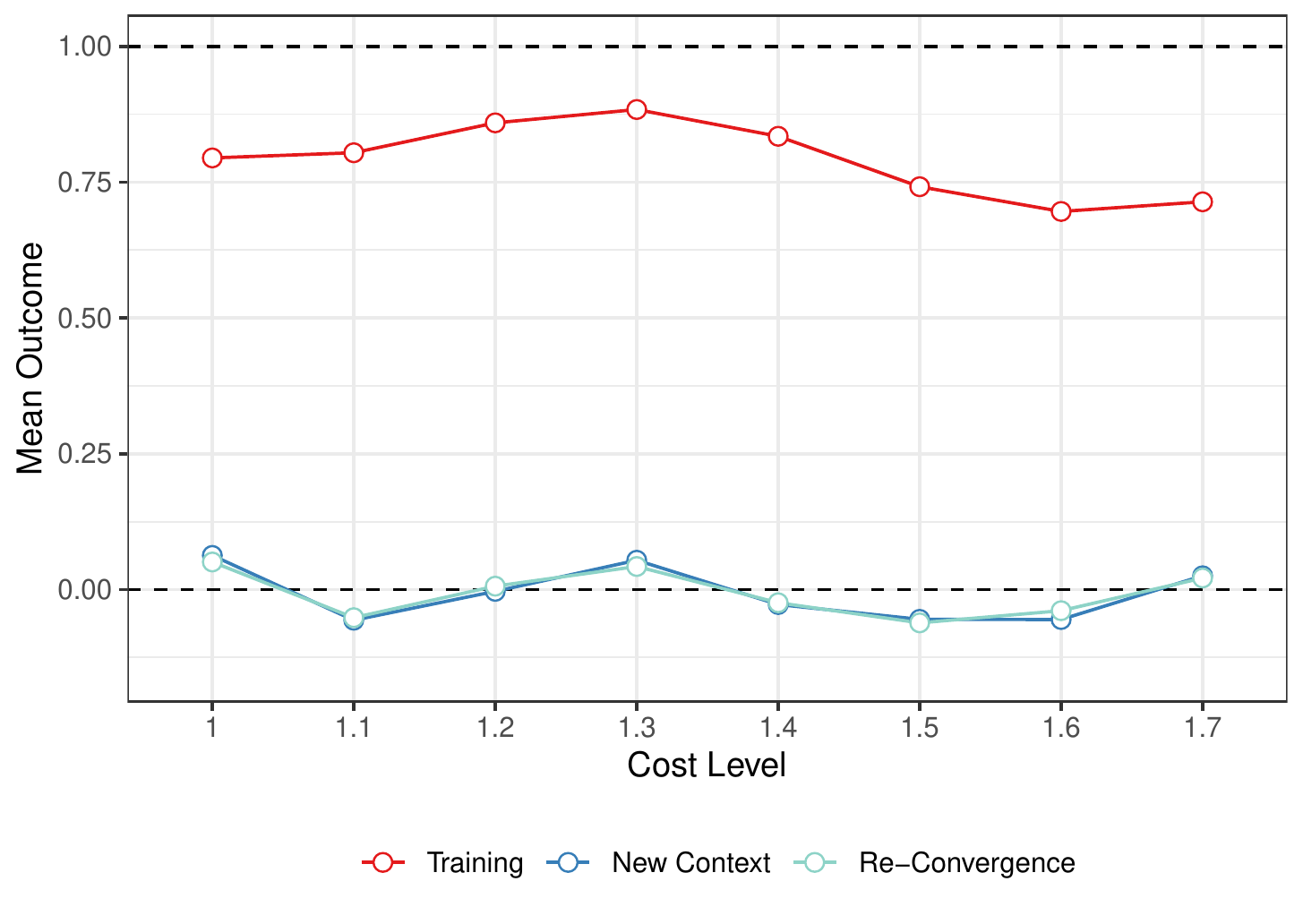}\\ \vspace{-1.5em}
  \flushleft{\footnotesize\textbf{Notes:} The figure shows the collusion index averaged across seeds in the (i) training contexts following convergence, (ii) first $1000$ iterations in the new testing contexts, and (iii) after convergence in the new contexts.}
  \label{fig:testing}
\end{figure}

\autoref{fig:testing} shows the average collusion index for each parameterization. In contrast to the outcome during training (in red), which shows the high collusion index detailed before, in the new context collusion breaks down. Instead, we observe Nash play on average. In order to exclude the possibility that collusion only vanishes due to the initial condition and it might simply take some iterations to restore it, we also let the algorithm run until policies converge again. This is captured by the lighter blue line (``Re-Convergence''). The result is the same: Irrespective of the number of periods of play, algorithmic collusion vanishes entirely.\footnote{We document the time to convergence during training and re-convergence, as well as the frequency of convergence types in \autoref{fig:convergencetime} and \autoref{fig:convergencetypes} in the Appendix.}

\begin{figure}[h]
  \centering
  \caption{ Collusion index in training and testing contexts with policy updating}
  \includegraphics[width=0.9\columnwidth]{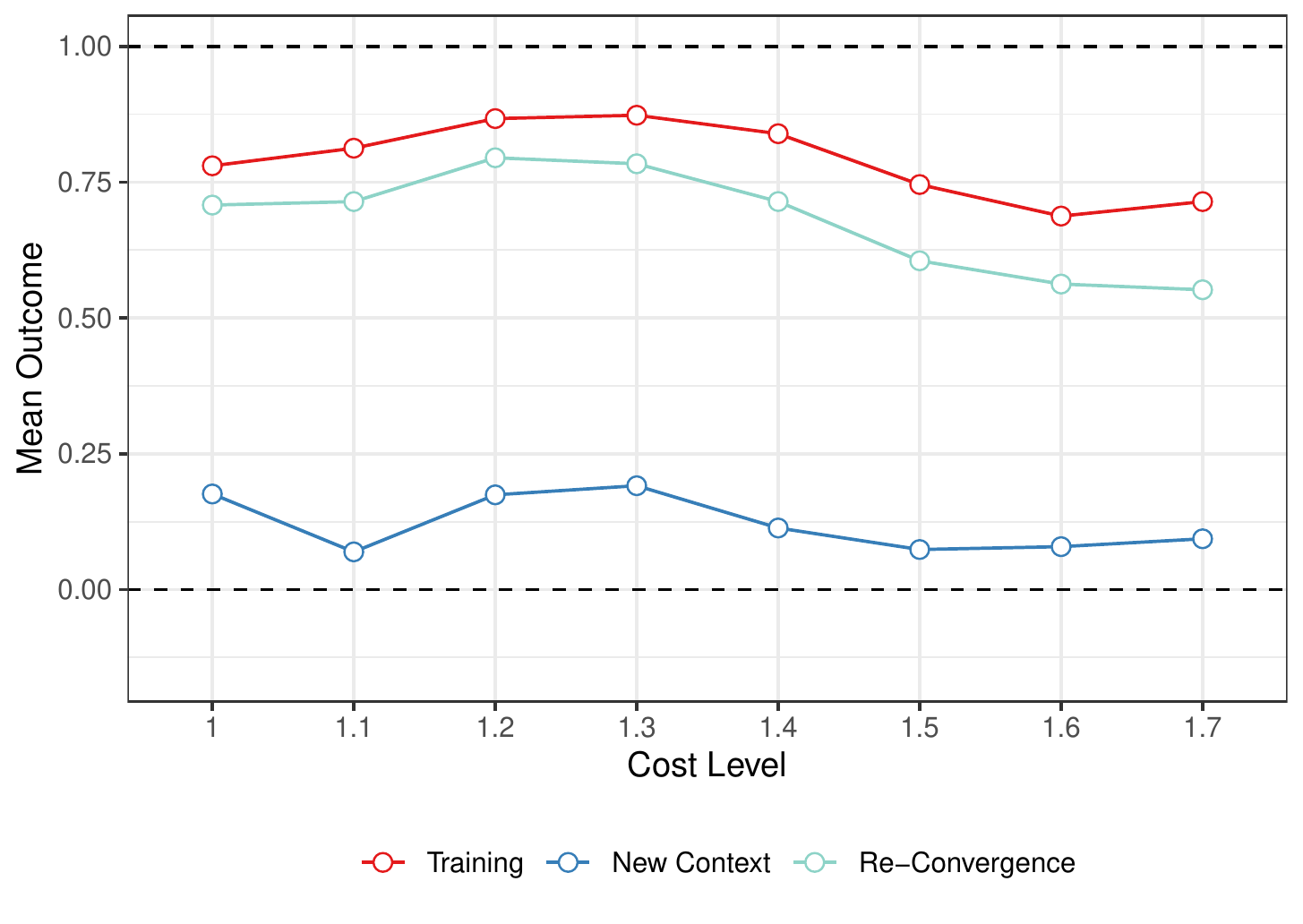}\\ \vspace{-1.5em}
  \flushleft{\footnotesize\textbf{Notes:} The figure shows the collusion index averaged across seeds when policies continue to be updated in the (i) training contexts following convergence, (ii) first $1000$ iterations in the new testing contexts, and (iii) after convergence in the new contexts.}
  \label{fig:updating}
\end{figure}

This finding shows that algorithmic collusion is not robust to changes in the environment, and performance during training is clearly no indication of outcome in the market. However, it seems likely that firms would not stop the learning process entirely in practice. As new information arrives while the algorithm is active, the policy can be updated further.  \autoref{fig:updating} displays the result when further reinforcing of the algorithms policy is enabled but exploration is not.

As before, in the new context collusion breaks down. However, we already observe profits on average that exceed Nash play. Once the algorithms have fully converged again, a relatively high level of collusion can be restored. This is in line with the literature that appears to find that algorithms achieve collusive outcomes in a large range of different environments and parameterizations \citep[][]{calvano2020, calvano2021varyingdemand, Klein2021}. Our analysis shows that this can potentially be achieved without exploration, if algorithms have been trained before in the ``correct'' environment. But it continues to take a long time for algorithms to converge, making it likely infeasible in practice (see \autoref{fig:convergenceupdating} in the Appendix).

We further study a policies' performance in a new context when the context differs in the parameterization of the environment. Suppose two pairs of algorithms train in contexts that differ in the marginal cost level. We assess their performance in a third context that is parameterized such that the parameters of each algorithm remain the same compared to their respective training context, but the parameters of the competing player are different in the testing context.

\begin{figure}[h]
\centering
  \caption{Collusion index in new contexts with different parameterization}
  \includegraphics[width=0.6\columnwidth]{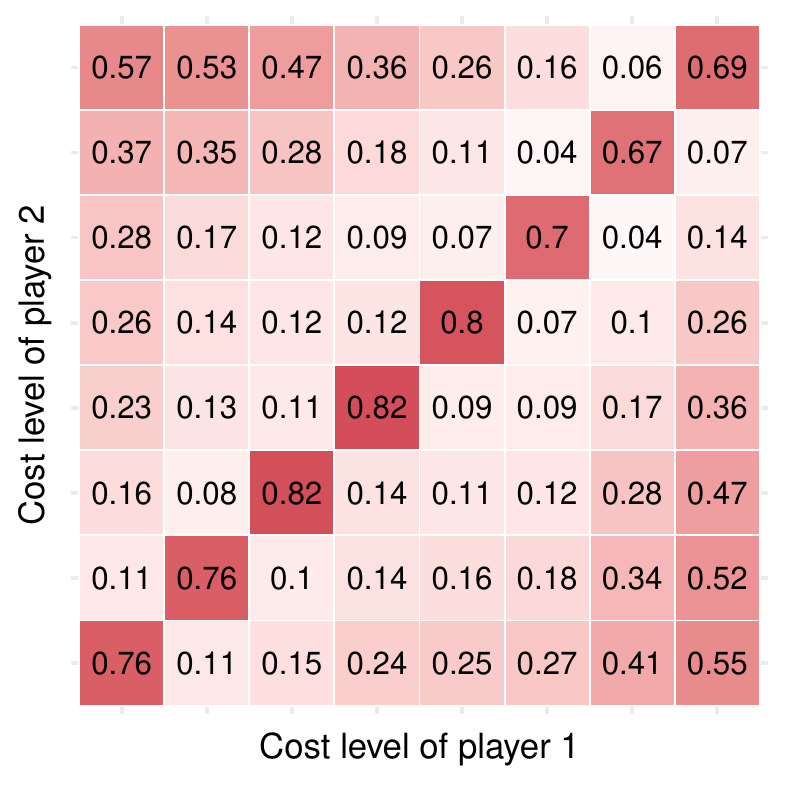}\\ \vspace{-1.5em}
  \flushleft{\footnotesize\textbf{Notes:} The figure shows the collusion index averaged across seeds in the training contexts following convergence (on-diagonal), and first $1000$ iterations in the new contexts (off-diagonal).}
  \label{fig:jpcci}
\end{figure}

\autoref{fig:jpcci} shows the result. Along the diagonal, we observe the training outcome (identical parameterization) documented before. Off-diagonal, we document the performance in asymmetric, differently-parameterized environments. We observe a breakdown of collusion and near-Nash play in particular for smaller parameter differences. We further quantify the effect of policies being employed in the new, asymmetric context (off-diagonal) by computing the \emph{average proportional loss} $R\_$, which is given by $R\_ = (\bar{D}-\bar{O})/\bar{D}$, where $\bar{D}$ is the mean value of the diagonal and $\bar{O}$ is the mean value of the off-diagonal entries. The average proportional loss for \autoref{fig:jpcci} is $R\_ = 0.72$. Thus, in testing contexts in which only the rivals' marginal cost level is different compared to the training context, the collusion index drops by 72\% on average. \autoref{fig:profitgainphase2} documents the average profit gain by player, showing that the higher values of the collusion index are driven by the relatively low-cost player achieving high profit gains. As before, collusion can be restored by continued learning (see \autoref{fig:jpcciphase3} and \autoref{fig:profitgainphase3} in the Appendix).

\subsection{Overfitting of Policies}

The breakdown of collusion we document is most severe for a change in context in which parameterization remains identical, or changes only marginally. We now examine the policies learnt during training to show that the driving force is overfitting to rival policies, and variation in ``off-equilibrium'' play.

\begin{figure}[h]
  \centering
  \caption{Two paths of play following deviation to Nash for intermediate parameterization with identical context except for the seed and identical convergence outcome during training}
  \begin{subfigure}{0.49\linewidth}
    \includegraphics[width=\linewidth]{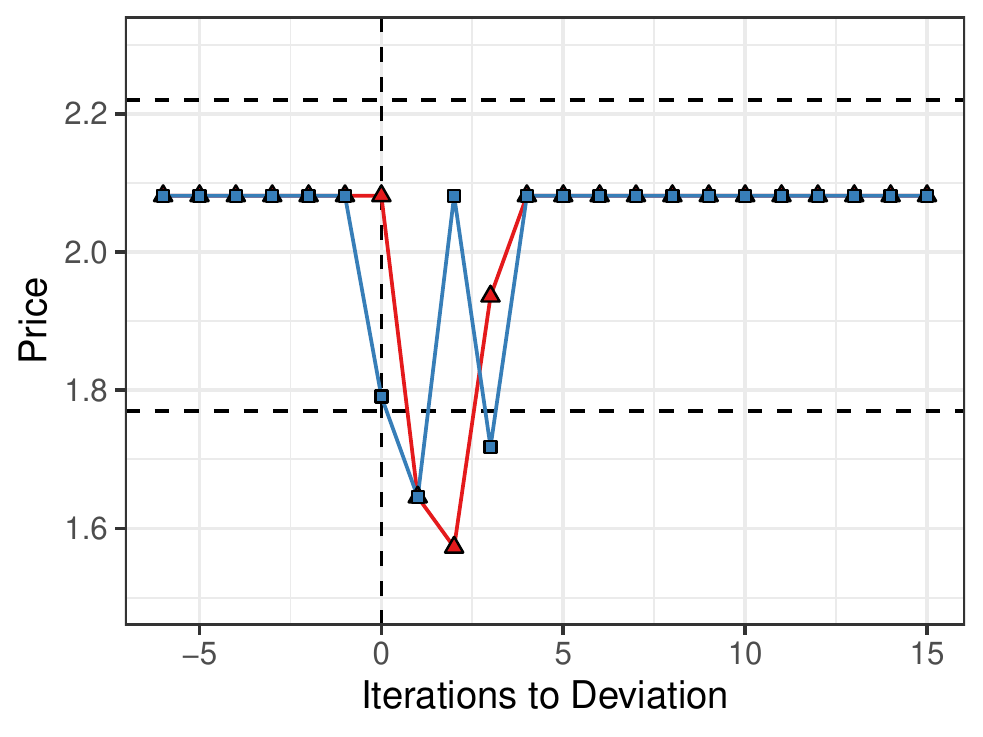}
  \end{subfigure}
  \begin{subfigure}{0.49\linewidth}
    \includegraphics[width=\linewidth]{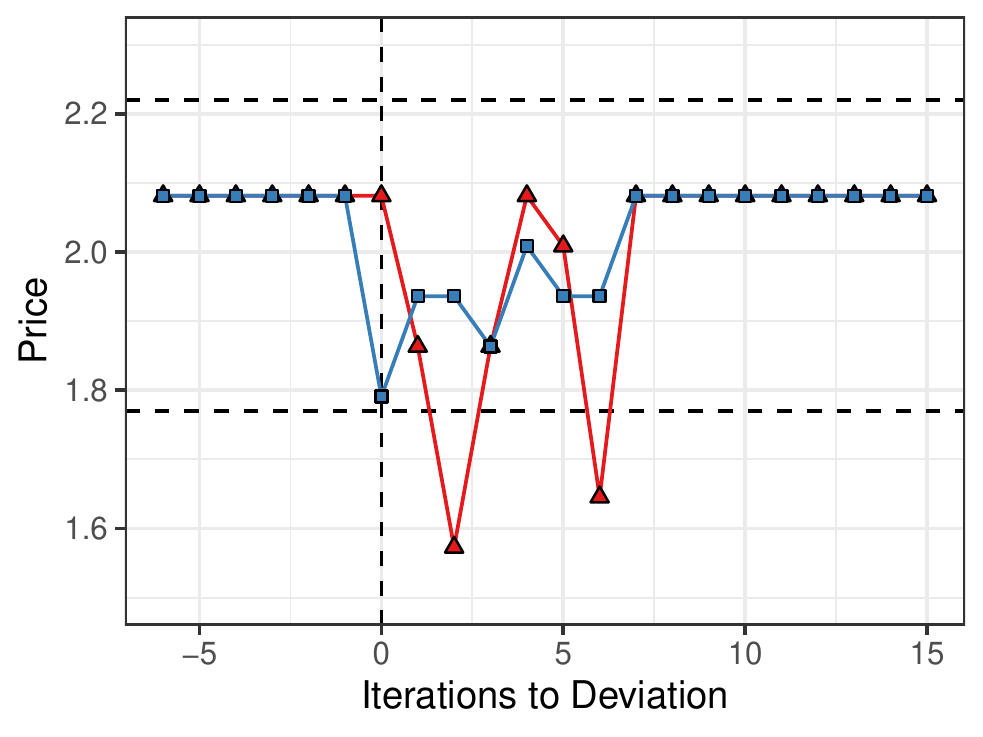}
  \end{subfigure}\\ \vspace{-0.5em}
  \flushleft{\footnotesize\textbf{Notes:} The figure shows the actions
    played prior to and following a manual deviation of one player
    ({\color{blue}{in blue}}) to the price closest to the Nash price on the grid
    for training contexts with intermediate cost parameterization for two
    different runs.
    Except for the initial seeds both runs use identical parameterizations and
    converged to identical outcomes during training.}
  \label{fig:twoexamplesdeviation}
\end{figure}

While we consistently observe ``reward-punishment'' patterns in converged play (see \autoref{fig:deviationpunishment}), the specific strategies differ greatly. For example, \autoref{fig:twoexamplesdeviation} shows the punishment following a deviation to the price nearest to the Nash price on the grid for two specific pairs of algorithms in two training contexts that solely differ in the seed used to initialize the learning process, and that converged to the same high, symmetric price pair.

To fully examine the variation in strategies played, we consider the transitive closure of strategy-pairs. Consider that once algorithms have converged, they are playing a pure-strategy that assigns one (optimal) action for each possible observation of the game for each player. We examine for a pair of jointly-converged policies during training and for each observation profile $o=(o_1,o_2)$ of the game the actions played according to the largest Q-value among all observation-action pairs for each player (i.e. the strategy-pair), and thus which new state they transition to.

\begin{figure}[h!]
  \centering
  \caption{Example strategy pair with multiple stable and unstable end-nodes}
  \includegraphics[width=0.9\columnwidth]{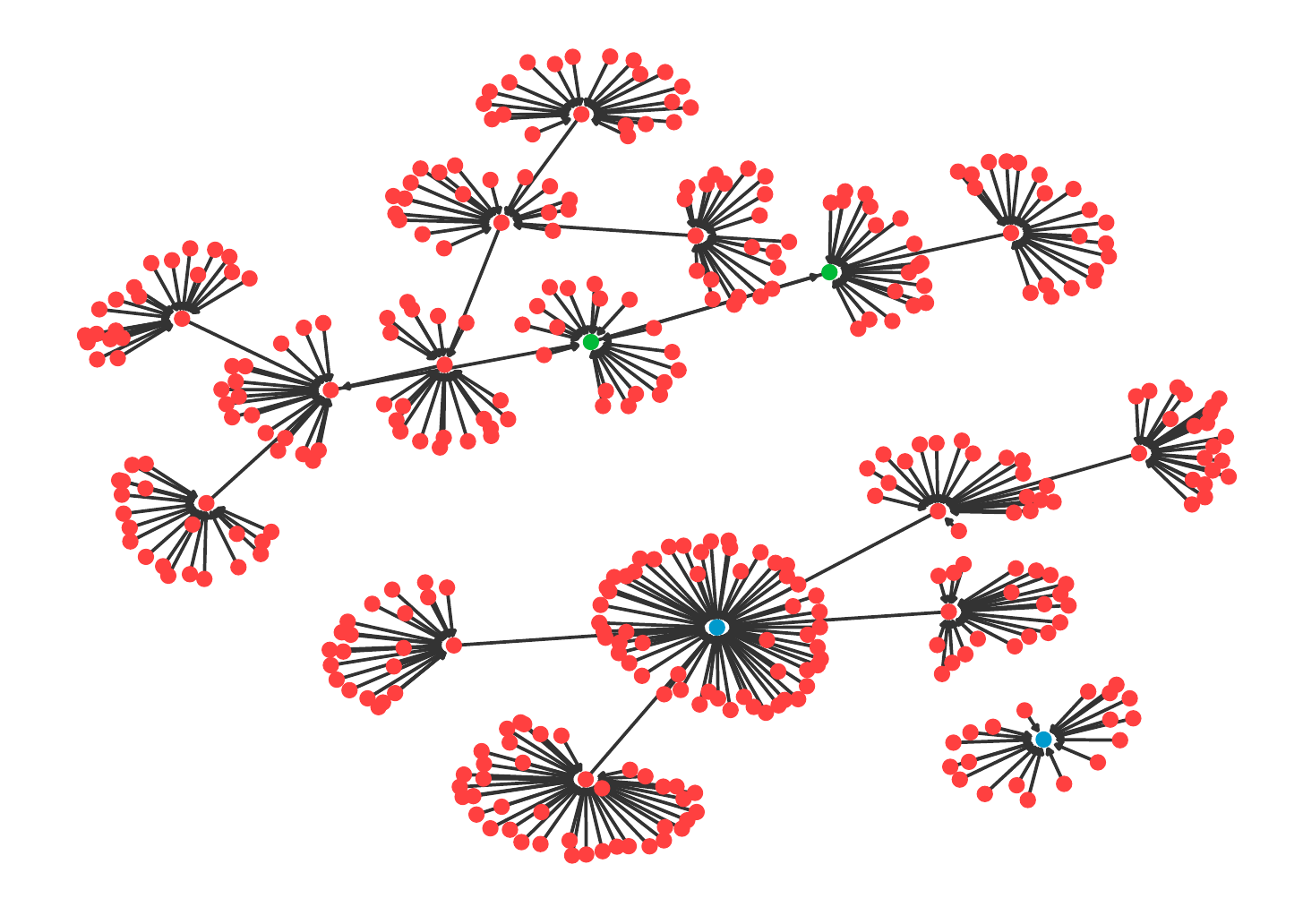}\\ \vspace{-1.5em}
  \flushleft{\footnotesize\textbf{Notes:} The figure shows the transitive closure of one pair of converged policies in their training context. Blue nodes are stable end-nodes, green nodes are unstable end-nodes.}
  \label{fig:network1}
\end{figure}

\autoref{fig:network1} illustrates the result for one pair of strategies of a
training context in which algorithms converged to a symmetric, high price level.
Each node is an observation profile and thus a state in the setting we consider, and the edges indicate the transitions that occur with this specific strategy-pair. Blue nodes are stable end-nodes, that is, once algorithms reach this action-pair, they will play it repeatedly forever. Green nodes are unstable end-nodes, meaning that a collection of neighboring green nodes are a cycle among action-pairs that will be played forever once algorithms reach it.

In the example shown, algorithms converged to a high, symmetric price, collusive outcome during training. This is the stable end-node at the center of \autoref{fig:network1}. Yet, depending on the state they find themselves in (i.e. actions played), the strategies do not necessarily lead back to this collusive outcome. That is, the outcome is not robust to perturbations. Moreover, when comparing strategy-pairs that each converged to \emph{the same} collusive outcome and the respective context differed solely in the seed used, we see significant variation. For example, \autoref{fig:network2} in the Appendix shows a strategy-pair that converged to the same collusive outcome as the one shown in \autoref{fig:network1}, but that in fact is robust to any perturbation: there is only one stable outcome.

\begin{figure}[h]
  \centering
  \caption{Two example strategy-pairs in testing context}
  \begin{subfigure}{0.49\linewidth}
  \includegraphics[width=\columnwidth]{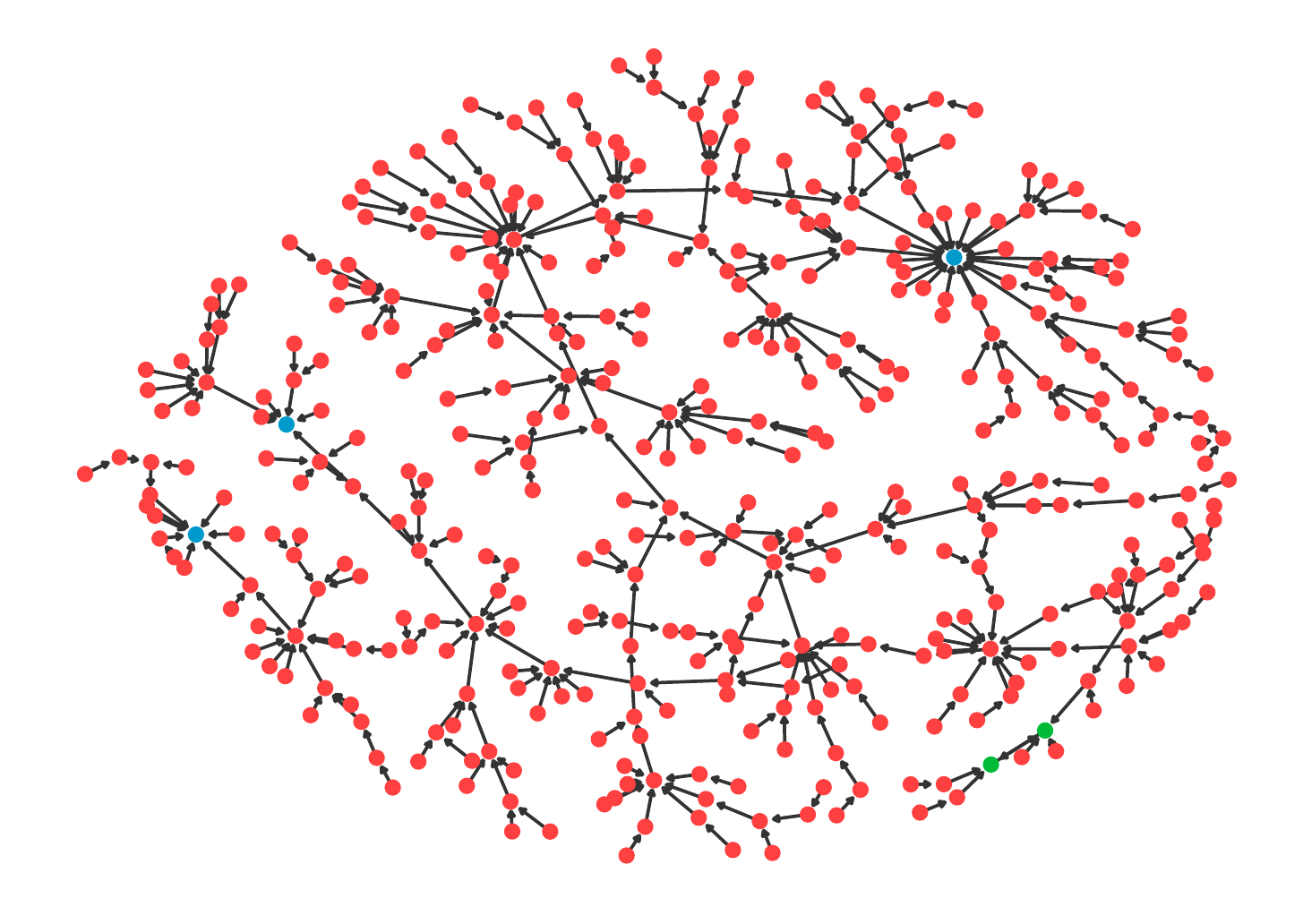}
  \end{subfigure}
  \begin{subfigure}{0.49\linewidth}
  \includegraphics[width=\columnwidth]{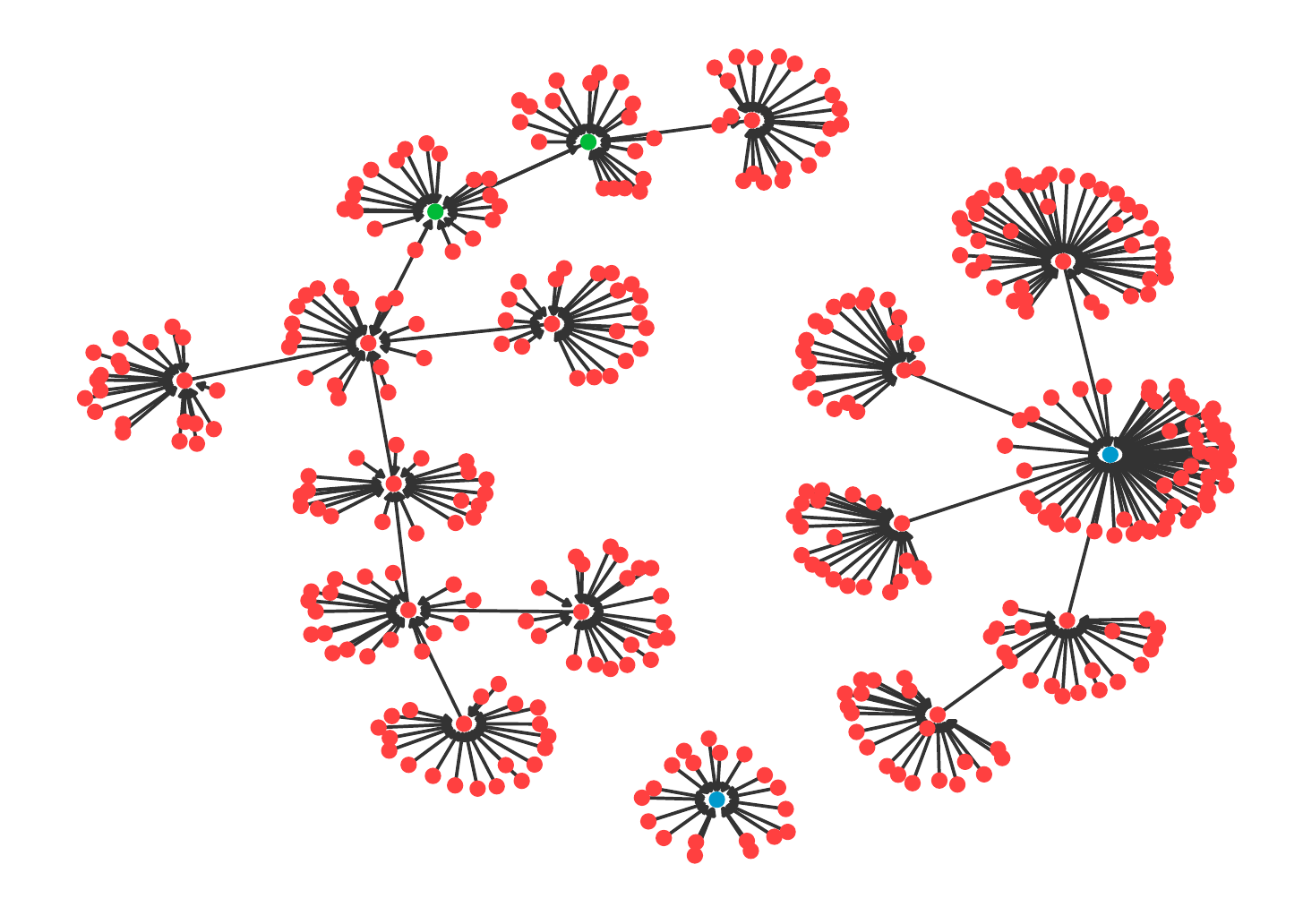}
\end{subfigure}\\ \vspace{-0.5em}
  \flushleft{\footnotesize\textbf{Notes:} The figure shows the transitive closure of two pairs of policies with identical outcome in training contexts evaluated in testing contexts. Blue nodes are stable end-nodes, green nodes are unstable end-nodes.}
  \label{fig:network3}
\end{figure}

When one strategy of each strategy pair is then placed in a third context (with identical parameterization), it is unsurprising that the collusive results from training cannot be extrapolated. \autoref{fig:network3} shows this for the specific strategy-pairs shown in \autoref{fig:network1} and \autoref{fig:network2}. In both panels, we place the strategy of one player from the first training context and one from the second jointly into a new, third context. We observe in both cases multiple possible outcomes, and none of them are the collusive outcome that both strategy-pairs played in their training context. The collusive policies that successfully established coordination in the training context are thus overfit to the specific policy of the competitor from training.

Viewed through the lens of standard game theory, the large variation in strategies that we observe is not necessarily surprising. When it comes to infinitely repeated games, the strategies which support an equilibrium  different from the static stage game Nash equilibrium are not unique \citep{mailath2006repeated}. Moreover, there is no guarantee that the learning agents in our setting learn equilibrium strategies perfectly. After all, they are only approximations.

\subsection{Robust Collusion}\label{subsec:robust}

The overfitting to rival policies that we document and the resulting instability of algorithmic collusion raises an important question: Is there a way for companies to design their learning algorithms so that they are more robust and achieve collusion in settings they have not been exposed to before? Our analysis shows that algorithms essentially learn to coordinate
on highly specific, overfit policies, which is only possible due to the large policy/strategy space. Hence, an obvious starting point is to restrict the algorithms' state space.\footnote{This is in line with research demonstrating that policies learnt via self-play can be exploited by adversarial policies particularly in high-dimensional environments, e.g. \cite{gleave2021adversarial}.}

\begin{figure}[h]
  \centering
  \caption{ Collusion index in training and testing contexts with restricted observation space}
  \includegraphics[width=0.9\columnwidth]{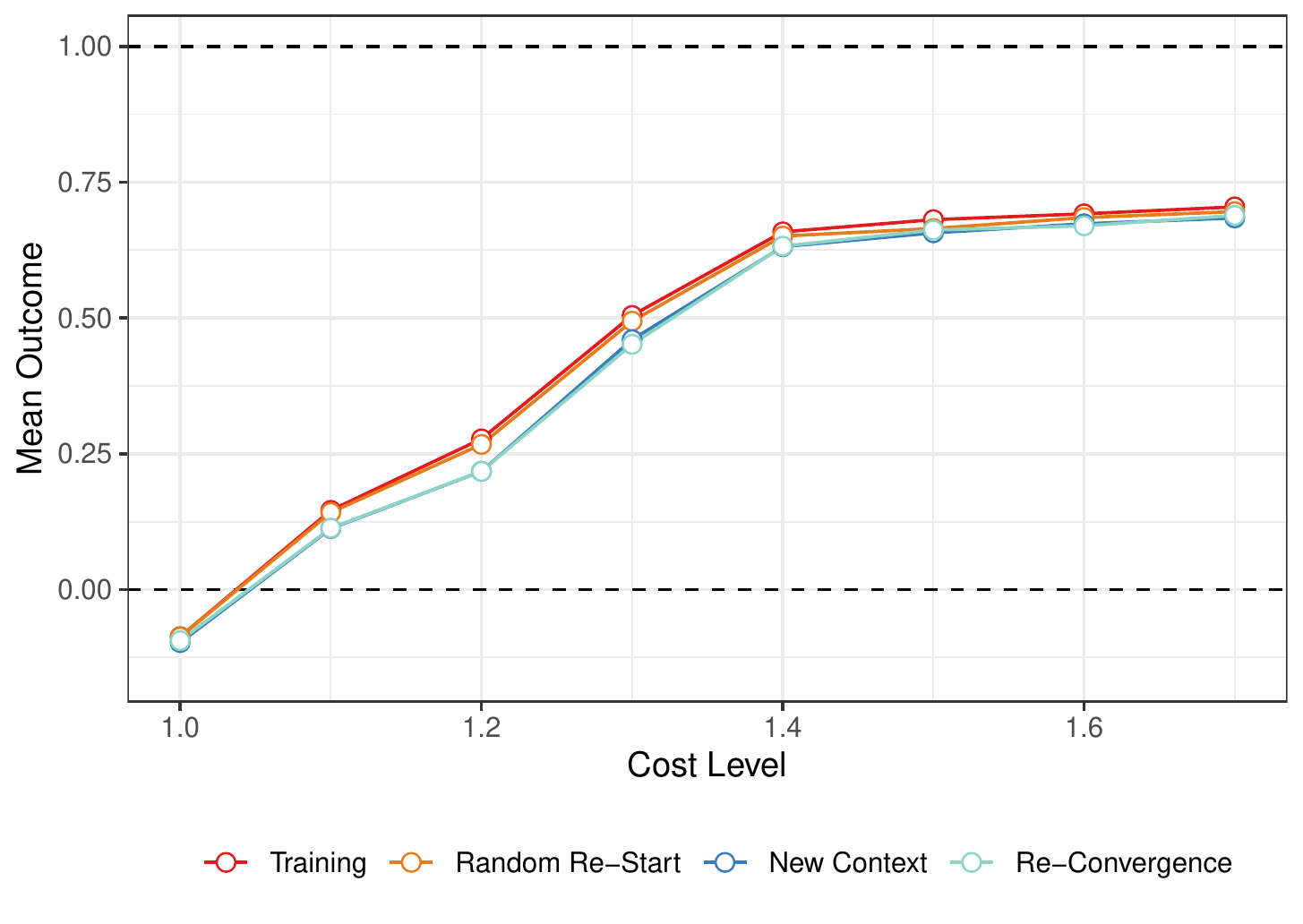}\\ \vspace{-1.5em}
  \flushleft{\footnotesize\textbf{Notes:} The figure shows the collusion index averaged across seeds in the (i) training contexts following convergence, (ii) first $1000$ iterations in training contexts with a random initial state, (iii) first $1000$ iterations in the new testing contexts, and (iv) after convergence in the new contexts, in the setting with a restricted observation space.}
  \label{fig:ownprice}
\end{figure}

We augment the setting introduced in \autoref{sec:model} by changing players' set of observations $O_i$. Specifically, algorithms now only observe their own price in the last period, so that $O = O_1 \times O_2 = (A_1 \times A_2)$ and $o_t = (o_{1,t} = p_{1,t-1}, o_{2,t} = p_{2, t-1})$.

\autoref{fig:ownprice} shows the collusion index averaged across contexts varying solely in their seed. We observe a high, seemingly-collusive outcome for parts of the differently-parameterized training contexts, specifically in higher cost contexts. Most importantly however, we observe that irrespective of the parameterization or level of the collusion index,  when placed in a new context the outcome is virtually identical to the training outcome. This does not change when we let algorithms play until converging again in the new context.

We provide an additional estimate in which we do not place algorithms in a new context, but place them in their training context with a random initial state. We do so to test explicitly that strategies are now robust to perturbations and generally yield a unique, stable outcome. \autoref{fig:ownprice} confirms this (``Random Re-Start''). Algorithms are now consistently able to extrapolate play from the training to the testing context.

\begin{figure}[h!]
  \centering
  \caption{Example strategy pair in new context}
  \includegraphics[width=0.9\columnwidth]{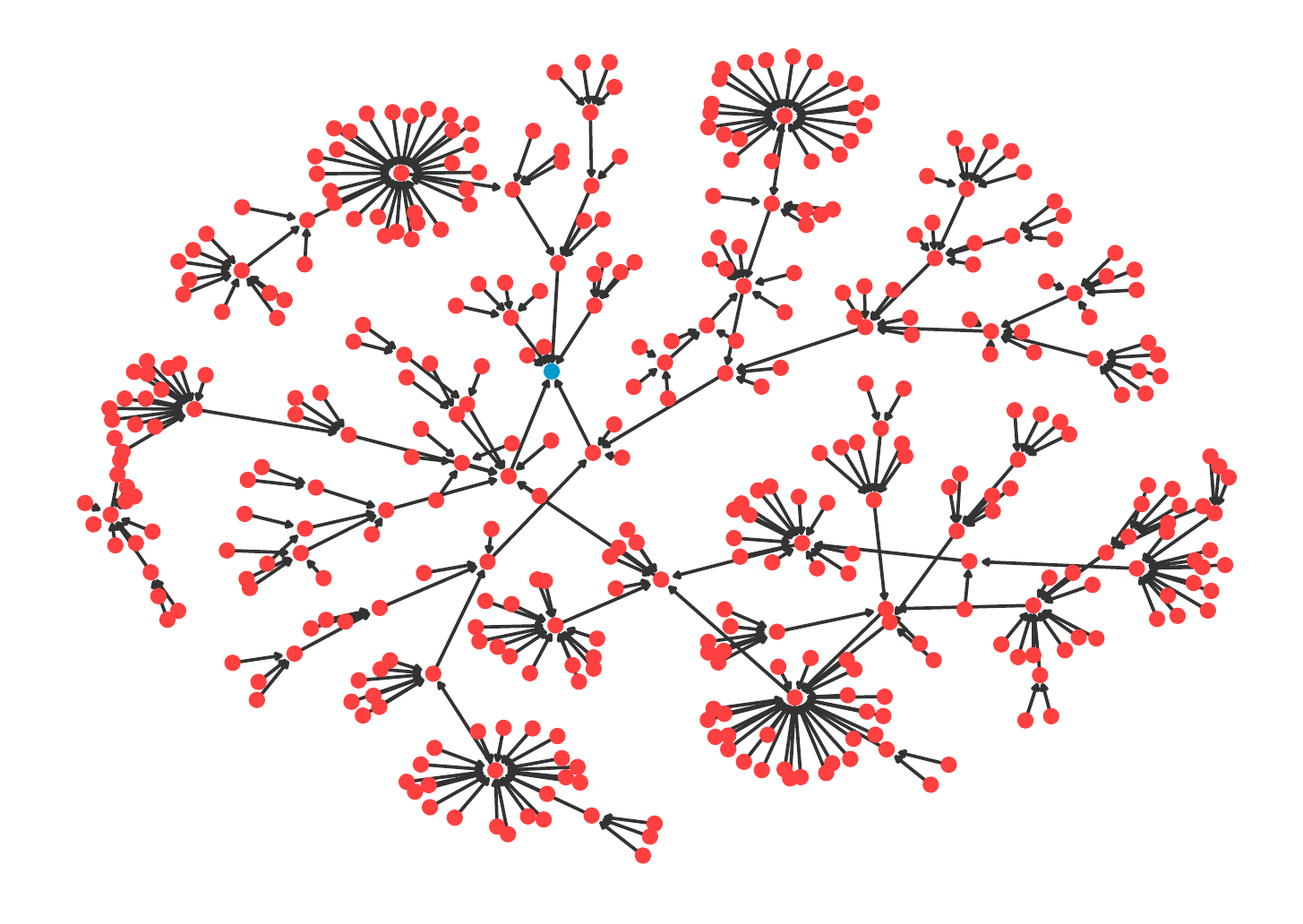}\\ \vspace{-1.5em}
  \flushleft{\footnotesize\textbf{Notes:} The figure shows the transitive closure of one pair of policies evaluated in the testing context for the setting with a restricted observation space. Blue nodes are stable end-nodes, green nodes are unstable end-nodes.}
  \label{fig:network4}
\end{figure}

\autoref{fig:network4} shows an example of a strategy-pair in the new (testing) context, that is typical for the strategy-pairs we observe. We now generally observe a unique, stable outcome even when playing against a previously unseen competitor. However, we also observe more short cycles compared to the baseline estimates from \autoref{subsec:learning} (we detail the frequency of convergence types in \autoref{fig:convergenceownpricetypes} in the Appendix).


\autoref{fig:ownpriceupdating} shows the outcomes when algorithms continue to update their policies in the new context. Interestingly, because policies from training are already robust to being placed in a new context, continuing to update policies is strictly worse than simply using the ``fixed'' policies from training.

\begin{figure}[h]
  \centering
  \caption{ Collusion index in training and testing contexts with restricted observation space and continued updating}
  \includegraphics[width=0.9\columnwidth]{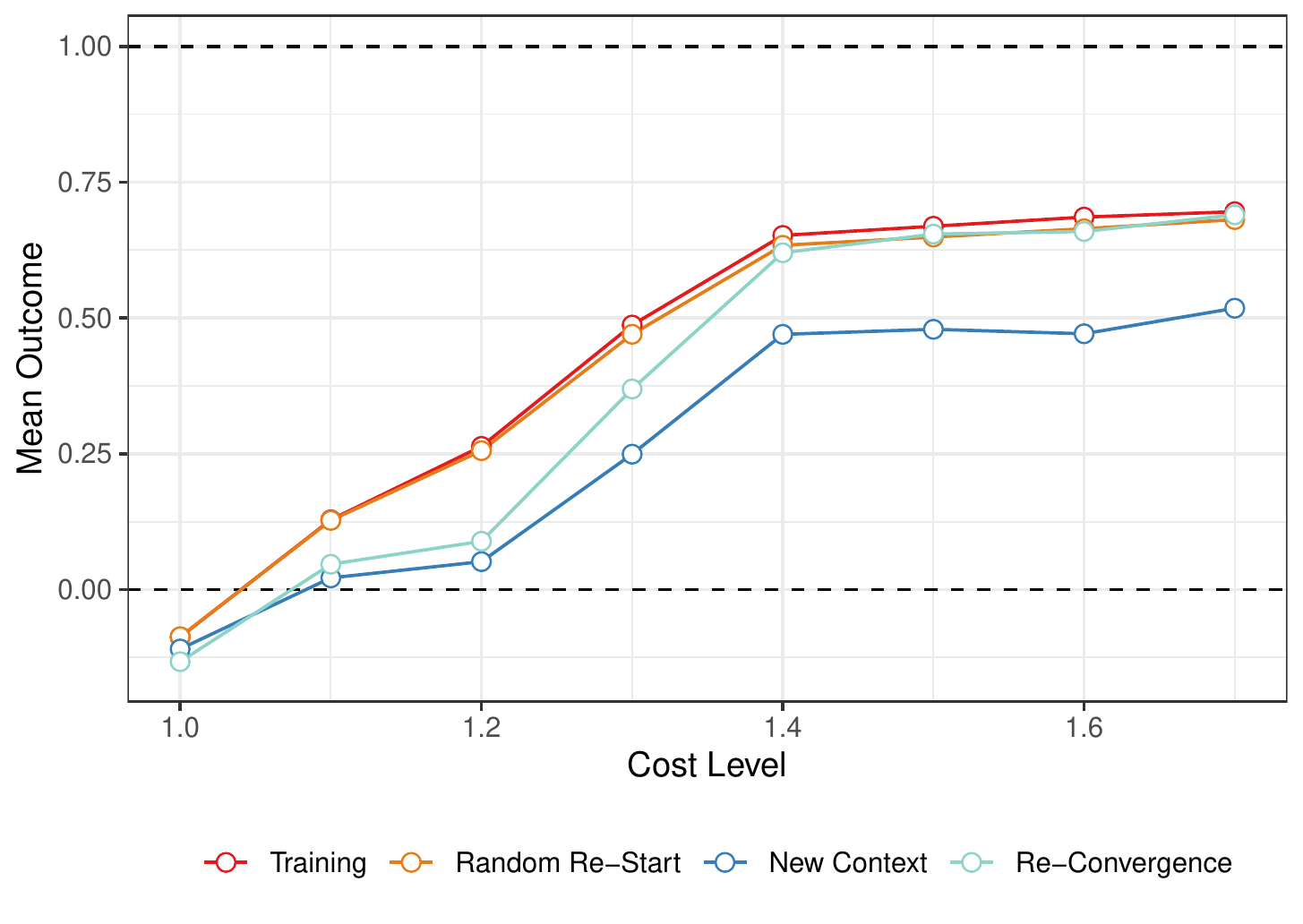}\\ \vspace{-1.5em}
  \flushleft{\footnotesize\textbf{Notes:} The figure shows the collusion index averaged across seeds when policies continue to be updated in the (i) training contexts following convergence, (ii) first $1000$ iterations in training contexts with a random initial state, (iii) first $1000$ iterations in the new testing contexts, and (iv) after convergence in the new contexts, in the setting with a restricted observation space.}
  \label{fig:ownpriceupdating}
\end{figure}

Our analysis suggests that a key driving force of algorithmic collusion in
practice is coordination of algorithm \emph{design}. Firms in practice may be
able to successfully achieve algorithmic (tacit) collusion by coordinating on
high-level ideas for the implementation of learning algorithms. Each firm then
implements and trains its algorithm independently. Yet by choosing the
parameterization of the environment, and the observation and thus policy space
appropriately, collusive policies can be made sufficiently robust to extrapolate to the market. Intuitively, precisely because the degree of coordination among competing firms is (legally) restricted, and algorithms must potentially work in a range of environments, the policy employed must rely on simpler patterns and cannot be too specific.\footnote{\cite{treutlein2021zeroshotcoordination} study the ``zero-shot coordination problem'' in a multiple settings and similarly note the importance for algorithms to ``only rely on general principles for coordination'' (p. 10).}

%

\section{Conclusion}\label{sec:conclusion}

Companies worldwide increasingly make use of reinforcement learning (RL) and other machine learning techniques for their pricing decisions. The application of such tools and the resulting automation of pricing decisions has gained the attention of policy-makers and researchers. One major concern is that self-learning pricing algorithms may lead to collusion without any explicit instruction to do so by firms. A nascent literature that studies the behavior of RL in economic games indicates that concerns about algorithmic collusion are not unfounded and algorithms can indeed learn sophisticated strategies supporting supra-competitive pricing. However, existing analyses have studied the behavior of algorithms in their training environment, but in practice the offline training environment is not identical to the market environment in which an algorithm is subsequently deployed.

This paper develops a framework to guide the analysis of learning algorithms in economic games. We consider a formal, general representation of the environment that is parameterized by a context. Evaluating policies learnt by algorithms in their training context in a suitably chosen testing context allows inference over the ability of policies to be extrapolated successfully to the market. We show that Q-learning algorithms in repeated Bertrand games overfit to rival strategies, and are unable to successfully use their collusive policies outside their training environment. In testing contexts, algorithmic collusion vanishes and does not recover with further iterations. Instead, we observe evidence of static Nash play, in particular with (nearly) identically parameterized contexts. We show that this is due to an overfit of policies to the rival training policy, and as a consequence we find that restricting algorithms' strategy space can make algorithmic collusion robust.

Our results provide a novel perspective on the existing findings of algorithmic collusion, and highlight the relevance of coordination at the level of algorithm design. While jointly learning algorithms appear prone to collusion, our analysis shows that this does not imply that collusive outcomes can be achieved in the market. Yet, by appropriately coordinating on aspects of the parameterization and strategy space, independently trained algorithms from rival firms may be designed to learn collusive policies that are robust to deployment in the market. Intuitively, because the extent of coordination among competing firms is (legally) restricted, and algorithms must work in a range of environments, the collusive policies employed must rely on simpler patterns and be fit to one another.

\clearpage
\bibliographystyle{apalike}
\bibliography{biblearning}

\begin{thebibliography}{}

\bibitem[Abrardi et~al., 2021]{Abrardi2021survey}
Abrardi, L., Cambini, C., and Rondi, L. (2021).
\newblock Artificial intelligence, firms and consumer behavior: A survey.
\newblock {\em Journal of Economic Surveys}, pages 1--23.

\bibitem[Acuna-Agost et~al., 2021]{acuna2021price}
Acuna-Agost, R., Thomas, E., and Lh{\'e}ritier, A. (2021).
\newblock Price elasticity estimation for deep learning-based choice models: an
  application to air itinerary choices.
\newblock {\em Journal of Revenue and Pricing Management}, 20(3):213--226.

\bibitem[Beneke and Mackenrodt, 2021]{beneke2021remedies}
Beneke, F. and Mackenrodt, M.-O. (2021).
\newblock Remedies for algorithmic tacit collusion.
\newblock {\em Journal of Antitrust Enforcement}, 9(1):152--176.

\bibitem[Bondoux et~al., 2020]{bondoux2020reinforcement}
Bondoux, N., Nguyen, A.~Q., Fiig, T., and Acuna-Agost, R. (2020).
\newblock Reinforcement learning applied to airline revenue management.
\newblock {\em Journal of Revenue and Pricing Management}, 19(5):332--348.

\bibitem[Calvano et~al., 2021]{calvano2021varyingdemand}
Calvano, E., Calzolari, G., Denicol{\`o}, V., and Pastorello, S. (2021).
\newblock Algorithmic collusion with imperfect monitoring.
\newblock {\em International Journal of Industrial Organization}, page 102712.

\bibitem[Calvano et~al., 2020a]{Calvano2020Science}
Calvano, E., Calzolari, G., Denicolò, V., Harrington, J.~E., and Pastorello,
  S. (2020a).
\newblock Protecting consumers from collusive prices due to ai.
\newblock {\em Science}, 370(6520):1040--1042.

\bibitem[Calvano et~al., 2020b]{calvano2020}
Calvano, E., Calzolari, G., Denicolò, V., and Pastorello, S. (2020b).
\newblock Artificial intelligence, algorithmic pricing, and collusion.
\newblock {\em American Economic Review}, 110(10):3267--97.

\bibitem[Gleave et~al., 2021]{gleave2021adversarial}
Gleave, A., Dennis, M., Wild, C., Kant, N., Levine, S., and Russell, S. (2021).
\newblock Adversarial policies: Attacking deep reinforcement learning.

\bibitem[Harrington, 2018]{harrington2018developing}
Harrington, J.~E. (2018).
\newblock Developing competition law for collusion by autonomous artificial
  agents.
\newblock {\em Journal of Competition Law \& Economics}, 14(3):331--363.

\bibitem[Harrington~Jr, 2021]{harrington2021effect}
Harrington~Jr, J.~E. (2021).
\newblock The effect of outsourcing pricing algorithms on market competition.
\newblock {\em Available at SSRN 3798847}.

\bibitem[Hu et~al., 2020]{hu2020otherplay}
Hu, H., Lerer, A., Peysakhovich, A., and Foerster, J. (2020).
\newblock ``other-play'' for zero-shot coordination.
\newblock In {\em International Conference on Machine Learning}, pages
  4399--4410. PMLR.

\bibitem[Justesen et~al., 2018]{justesen2018illuminating}
Justesen, N., Torrado, R.~R., Bontrager, P., Khalifa, A., Togelius, J., and
  Risi, S. (2018).
\newblock Illuminating generalization in deep reinforcement learning through
  procedural level generation.

\bibitem[Kastius and Schlosser, 2021]{kastius2021dynamic}
Kastius, A. and Schlosser, R. (2021).
\newblock Dynamic pricing under competition using reinforcement learning.
\newblock {\em Journal of Revenue and Pricing Management}, pages 1--14.

\bibitem[Kirk et~al., 2021]{kirkSurveyGeneralisationDeep2021}
Kirk, R., Zhang, A., Grefenstette, E., and Rockt{\"a}schel, T. (2021).
\newblock A {{Survey}} of {{Generalisation}} in {{Deep Reinforcement
  Learning}}.
\newblock {\em arXiv:2111.09794 [cs]}.

\bibitem[Klein, 2021]{Klein2021}
Klein, T. (2021).
\newblock Autonomous algorithmic collusion: Q-learning under sequential
  pricing.
\newblock {\em The RAND Journal of Economics}, 52(3):538--558.

\bibitem[Lanctot et~al., 2017]{lanctot2017unified}
Lanctot, M., Zambaldi, V., Gruslys, A., Lazaridou, A., Tuyls, K., P{\'e}rolat,
  J., Silver, D., and Graepel, T. (2017).
\newblock A unified game-theoretic approach to multiagent reinforcement
  learning.
\newblock {\em arXiv preprint arXiv:1711.00832}.

\bibitem[Mailath and Samuelson, 2006]{mailath2006repeated}
Mailath, G.~J. and Samuelson, L. (2006).
\newblock {\em Repeated games and reputations: long-run relationships}.
\newblock Oxford university press.

\bibitem[Nichol et~al., 2018]{nichol2018gotta}
Nichol, A., Pfau, V., Hesse, C., Klimov, O., and Schulman, J. (2018).
\newblock Gotta learn fast: A new benchmark for generalization in rl.
\newblock {\em arXiv preprint arXiv:1804.03720}.

\bibitem[Ohlhausen, 2017]{ohlhausen2017lawalgos}
Ohlhausen, M.~K. (2017).
\newblock Should we fear the things that go beep in the night? some initial
  thoughts on the intersection of antitrust law and algorithmic pricing.

\bibitem[Song et~al., 2019]{song2019overfitting}
Song, X., Jiang, Y., Tu, S., Du, Y., and Neyshabur, B. (2019).
\newblock Observational overfitting in reinforcement learning.
\newblock {\em arXiv preprint arXiv:1912.02975}.

\bibitem[Treutlein et~al., 2021]{treutlein2021zeroshotcoordination}
Treutlein, J., Dennis, M., Oesterheld, C., and Foerster, J. (2021).
\newblock A new formalism, method and open issues for zero-shot coordination.

\bibitem[Vestager, 2017]{vestager2017algosberlin}
Vestager, M. (2017).
\newblock Algorithms and competition.
\newblock In {\em Bundeskartellamt 18th Conference on Competition, Berlin, 16
  March 2017}.

\bibitem[Zhang et~al., 2018a]{zhang2018overfittingcontinuous}
Zhang, A., Ballas, N., and Pineau, J. (2018a).
\newblock A dissection of overfitting and generalization in continuous
  reinforcement learning.
\newblock {\em arXiv preprint arXiv:1806.07937}.

\bibitem[Zhang et~al., 2018b]{zhang2018overfitting}
Zhang, C., Vinyals, O., Munos, R., and Bengio, S. (2018b).
\newblock A study on overfitting in deep reinforcement learning.
\newblock {\em arXiv preprint arXiv:1804.06893}.

\end{thebibliography}

\clearpage
\newpage
\appendix
\appendixpage
\section{Figures}
\setcounter{figure}{0}
\renewcommand{\thefigure}{\Alph{section}\arabic{figure}}

\begin{figure}[h]
  \centering
  \caption{Time to convergence without continued policy updating}
  \label{fig:convergencetime}
  \begin{subfigure}{0.495\textwidth}
    \caption{Training}
    \includegraphics[width=\linewidth]{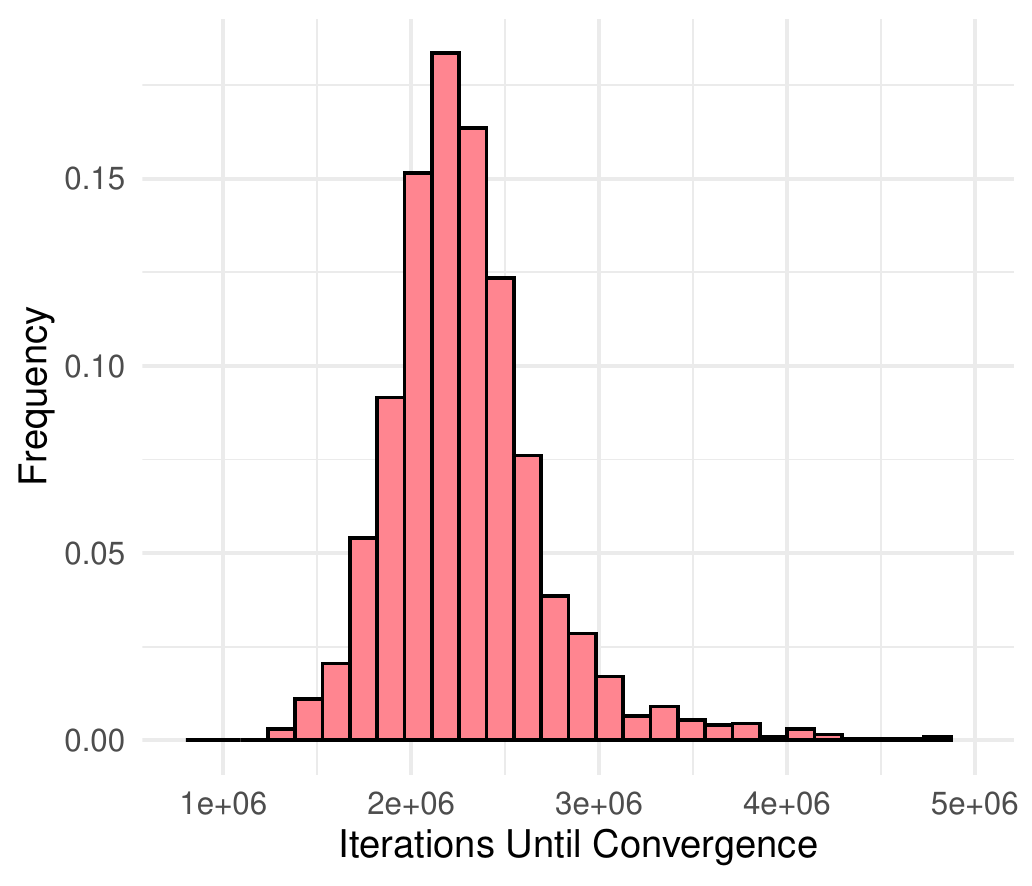}
  \end{subfigure}
  \begin{subfigure}{0.495\textwidth}
    \caption{Re-convergence}
    \includegraphics[width=\linewidth]{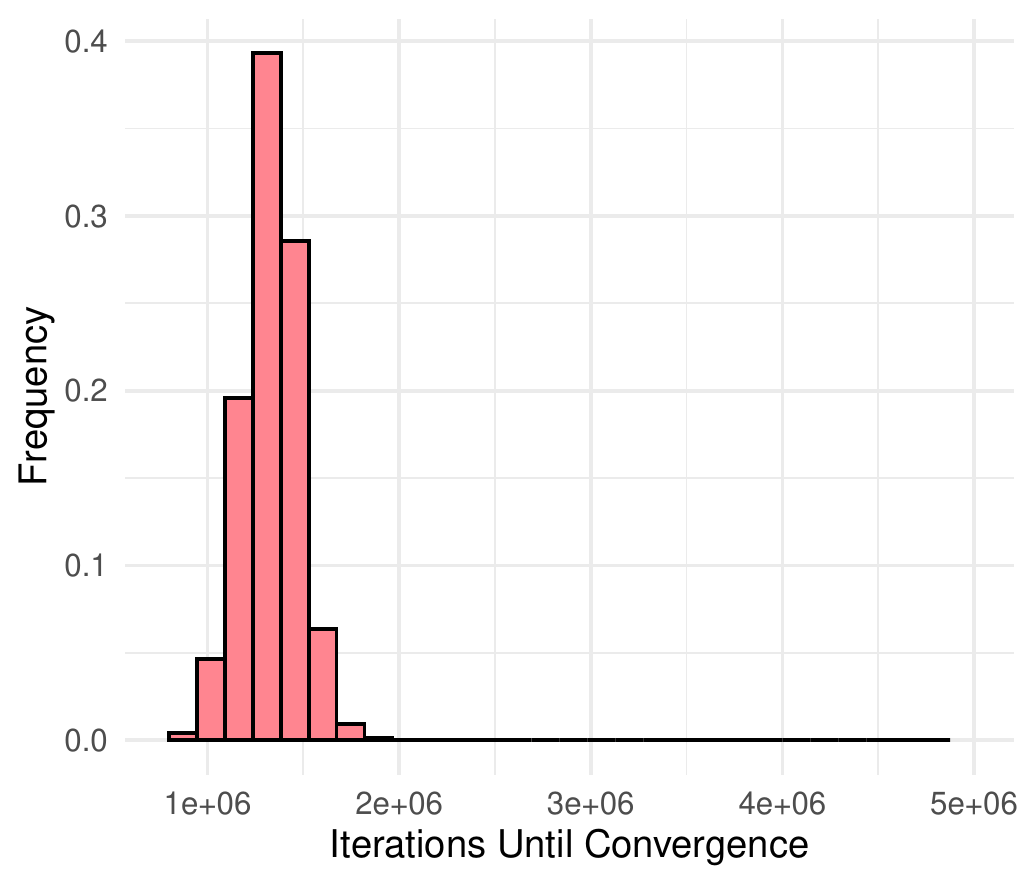}
  \end{subfigure}\\ \vspace{-0.5em}
  \flushleft{\footnotesize\textbf{Notes:} The figure shows a histogram of the number of iterations until policies converged in the training contexts (panel (a)) and testing contexts with identical parameterization (panel (b)).}
\end{figure}

\begin{figure}
  \centering
  \caption{Convergence types without continued policy updating}
  \label{fig:convergencetypes}
  \begin{subfigure}{0.495\textwidth}
    \caption{Training}
    \includegraphics[width=\linewidth]{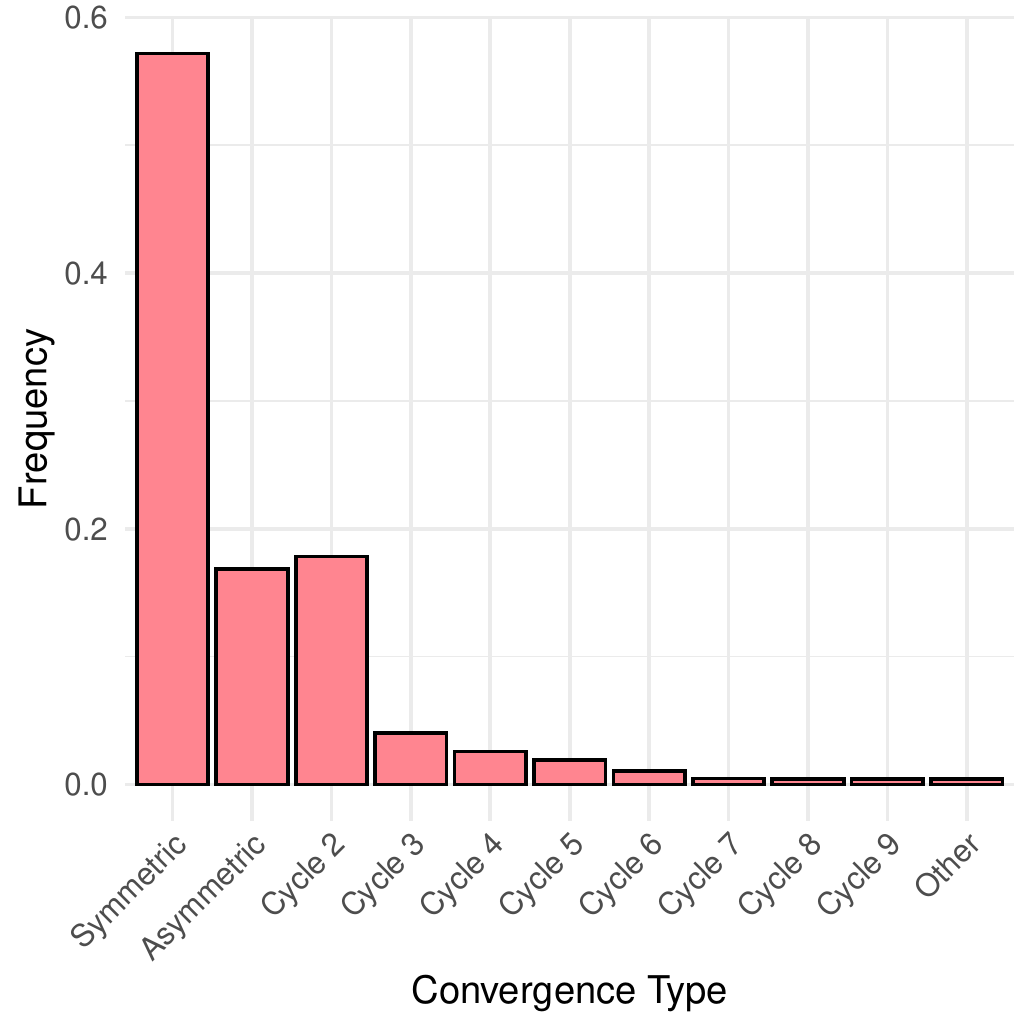}
  \end{subfigure}
  \begin{subfigure}{0.495\textwidth}
    \caption{Re-convergence}
    \includegraphics[width=\linewidth]{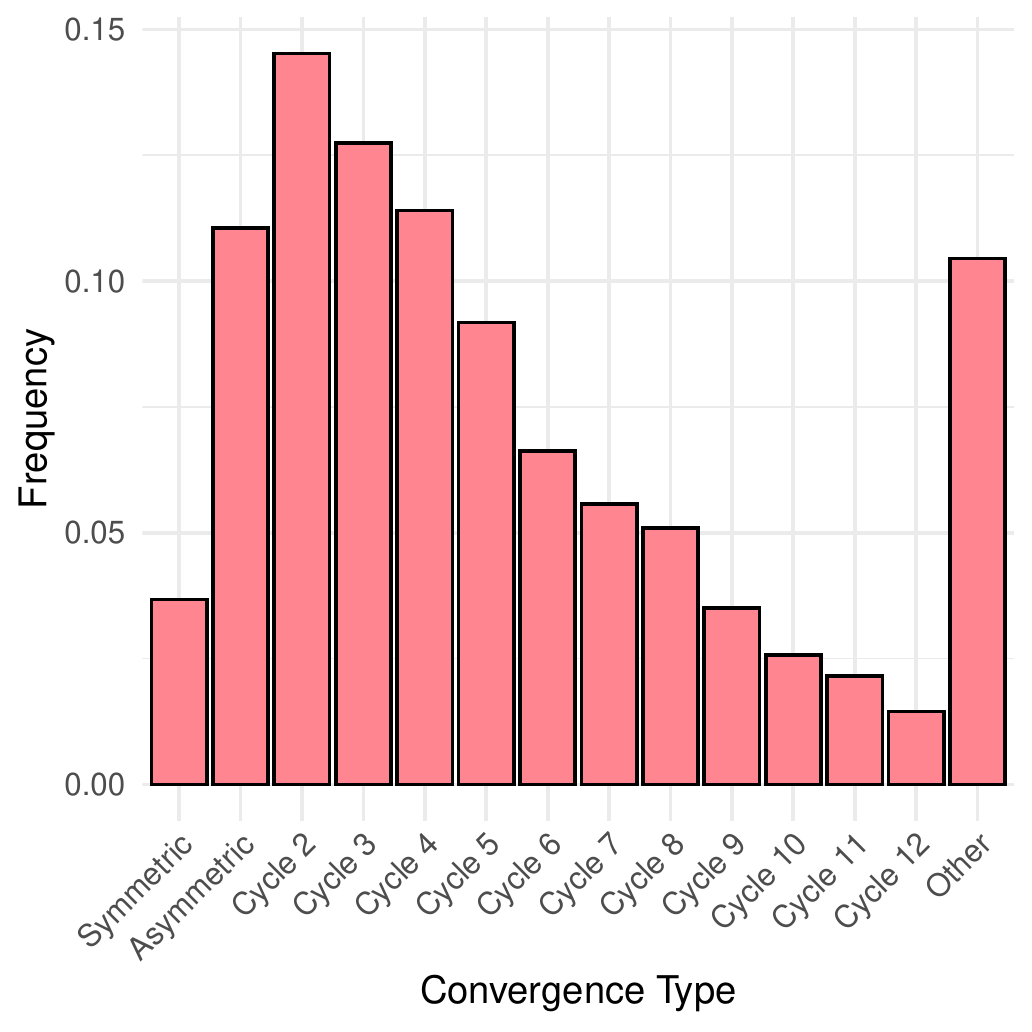}
  \end{subfigure}\\ \vspace{-0.5em}
  \flushleft{\footnotesize\textbf{Notes:} The figure shows a histogram of the frequency of convergence types in training contexts (panel (a)) and testing contexts with identical parameterization (panel (b)).}
\end{figure}

\begin{figure}[h]
\centering
\caption{Time to re-convergence with continued policy updating}
\includegraphics[width = 0.6\columnwidth]{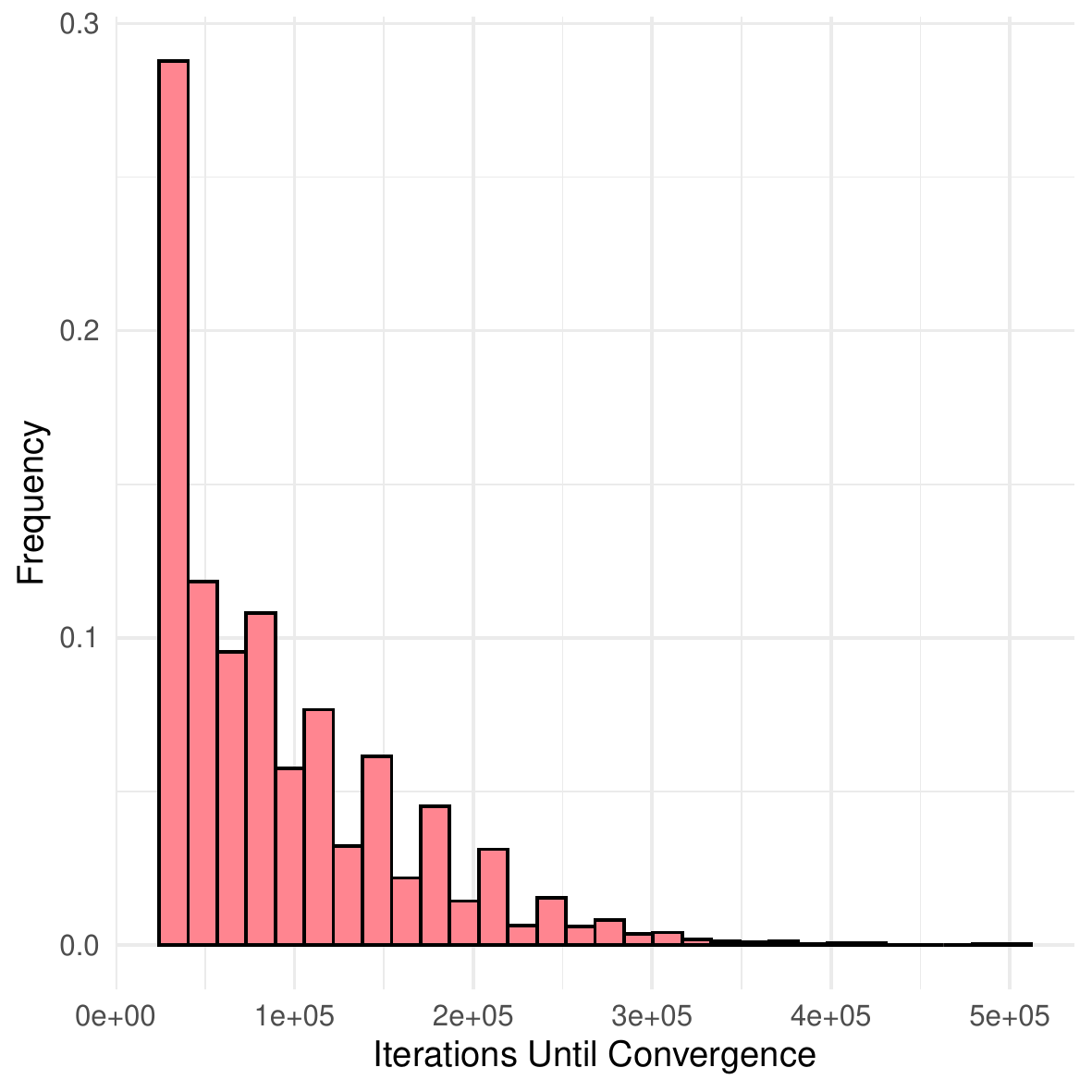}\\ \vspace{-0.5em}
\flushleft{\footnotesize\textbf{Notes:} The figure shows a histogram of the number of iterations until policies converged in the training contexts (panel (a)) and testing contexts with identical parameterization (panel (b)) when policies continue to be updated.}
\label{fig:convergenceupdating}
\end{figure}

\begin{figure}[h]
\centering
\caption{Convergence types with continued updating}
\begin{subfigure}{0.49\columnwidth}
\subcaption{Training}
\includegraphics[width = \columnwidth]{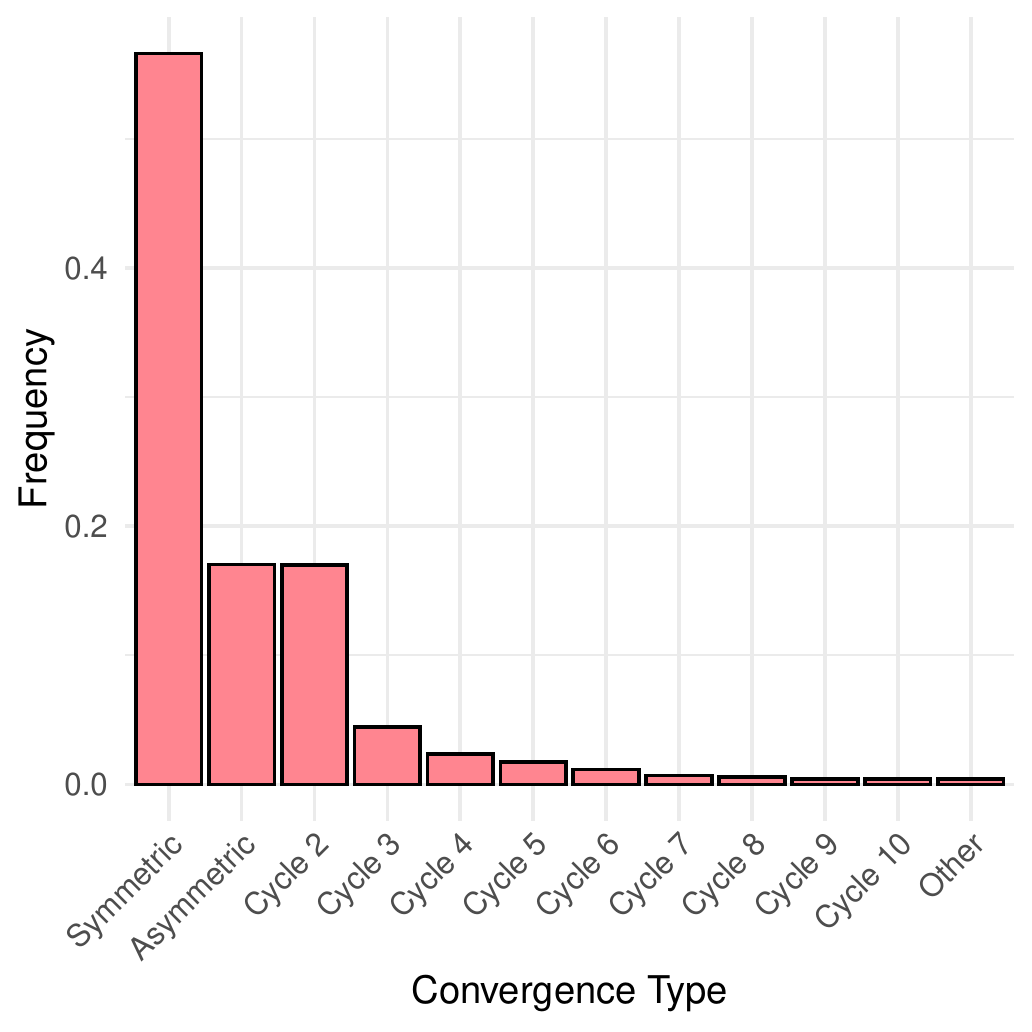}
\end{subfigure}
\begin{subfigure}{0.49\columnwidth}
\subcaption{Re-convergence}
\includegraphics[width = \columnwidth]{img/asymmetricLearners3PhasesRobustnessExplorationConvergenceLearningAllCosts-242ad22745c25c1e47979b09bf7f621a786c1a78/equilibria_frequency_phase1.pdf}
\end{subfigure}\\ \vspace{-0.5em}
\flushleft{\footnotesize\textbf{Notes:} The figure shows a histogram of the frequency of convergence types in training contexts (panel (a)) and testing contexts with identical parameterization (panel (b)) when policies continue to be updated.}
\label{fig:convergenceupdatingtype}
\end{figure}

\begin{figure}[h]
  \centering
  \caption{Profit gain by player}
  \label{fig:profitgainphase2}
  \includegraphics[width=\linewidth]{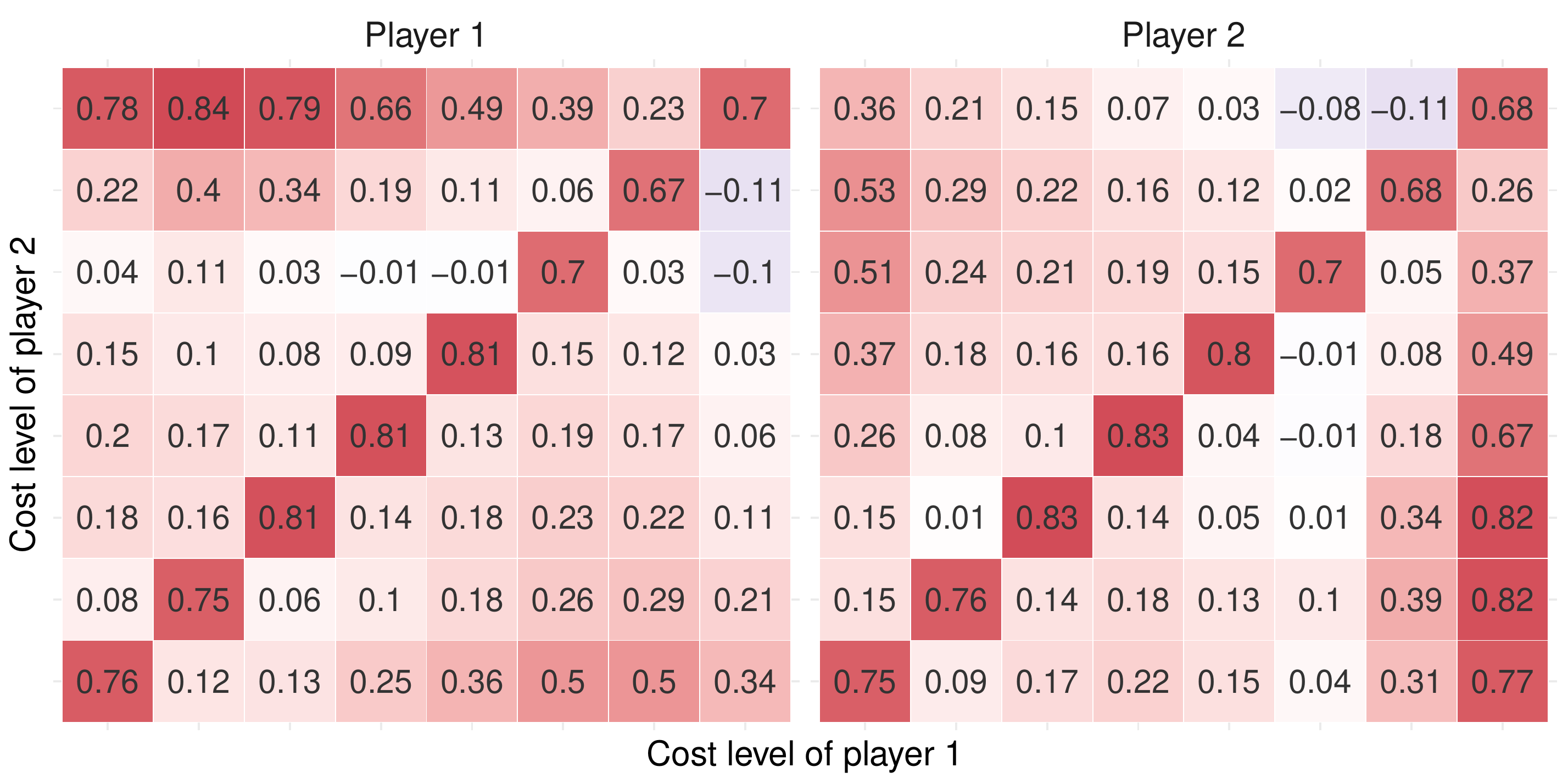}\\ \vspace{-1.5em}
  \flushleft{\footnotesize\textbf{Notes:} The figure shows the profit gain by player averaged across seeds for all testing contexts with asymmetric parameterizations.}
\end{figure}

\begin{figure}[h]
\centering
\caption{Collusion index after re-convergence in testing contexts}
\includegraphics[width = 0.6\columnwidth]{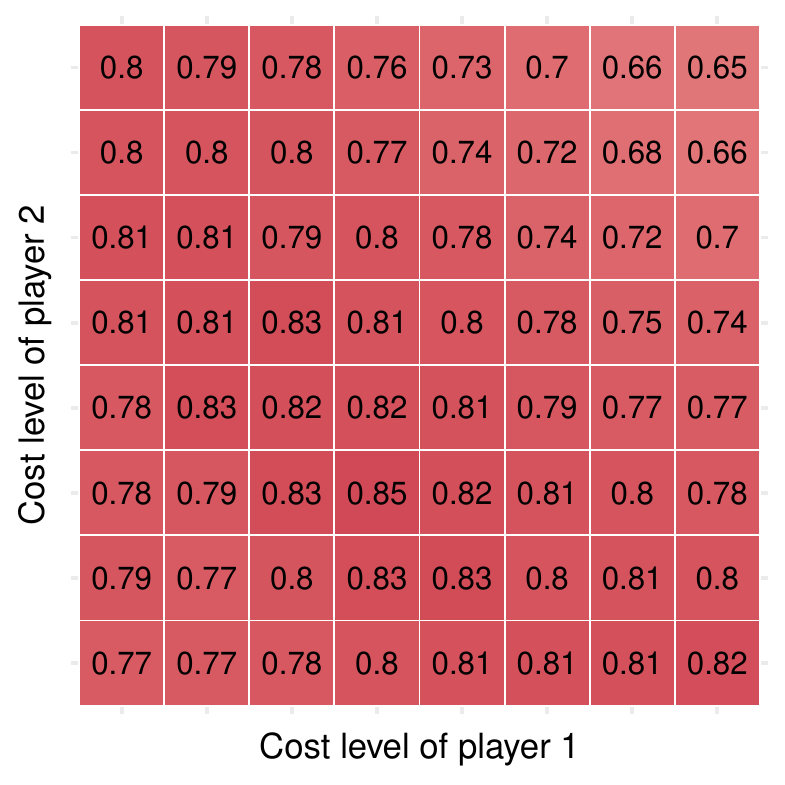}\\ \vspace{-1.5em}
\flushleft{\footnotesize\textbf{Notes:} The figure shows the collusion index averaged across seeds for all testing contexts with asymmetric parameterizations following re-convergence.}
\label{fig:jpcciphase3}
\end{figure}

\begin{figure}[h]
  \centering
  \caption{Profit gain by player after re-convergence}
  \label{fig:profitgainphase3}
  \includegraphics[width=\linewidth]{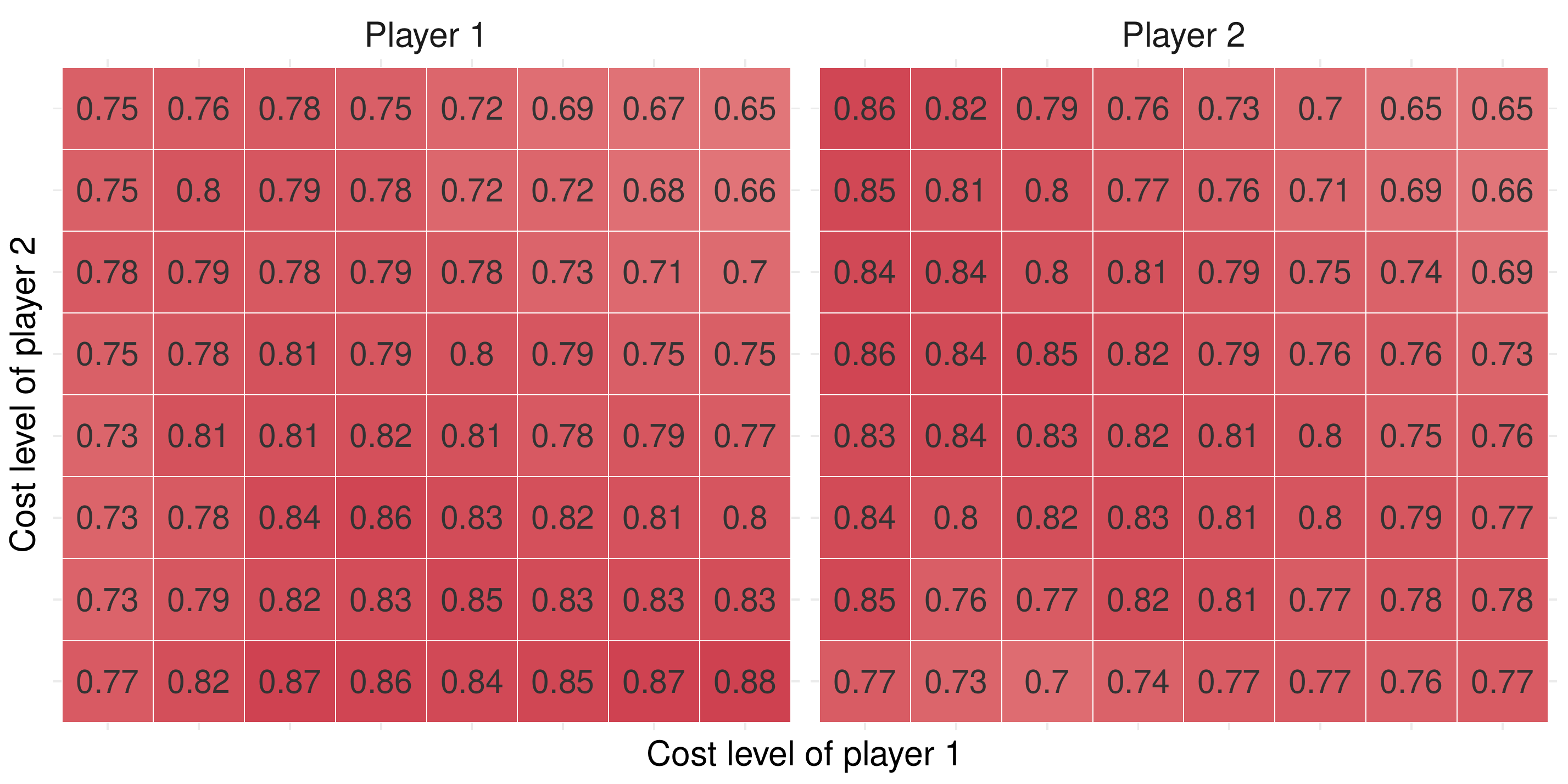}\\ \vspace{-1.5em}
  \flushleft{\footnotesize\textbf{Notes:} The figure shows the profit gain by player averaged across seeds for all testing contexts with asymmetric parameterizations following re-convergence.}
\end{figure}

\begin{figure}
  \centering
  \caption{Convergence types with asymmetric parameterized testing contexts}
  \label{fig:convergencetype}
  \begin{subfigure}{0.495\textwidth}
    \caption{On-diagonal training context}
    \includegraphics[width=\linewidth]{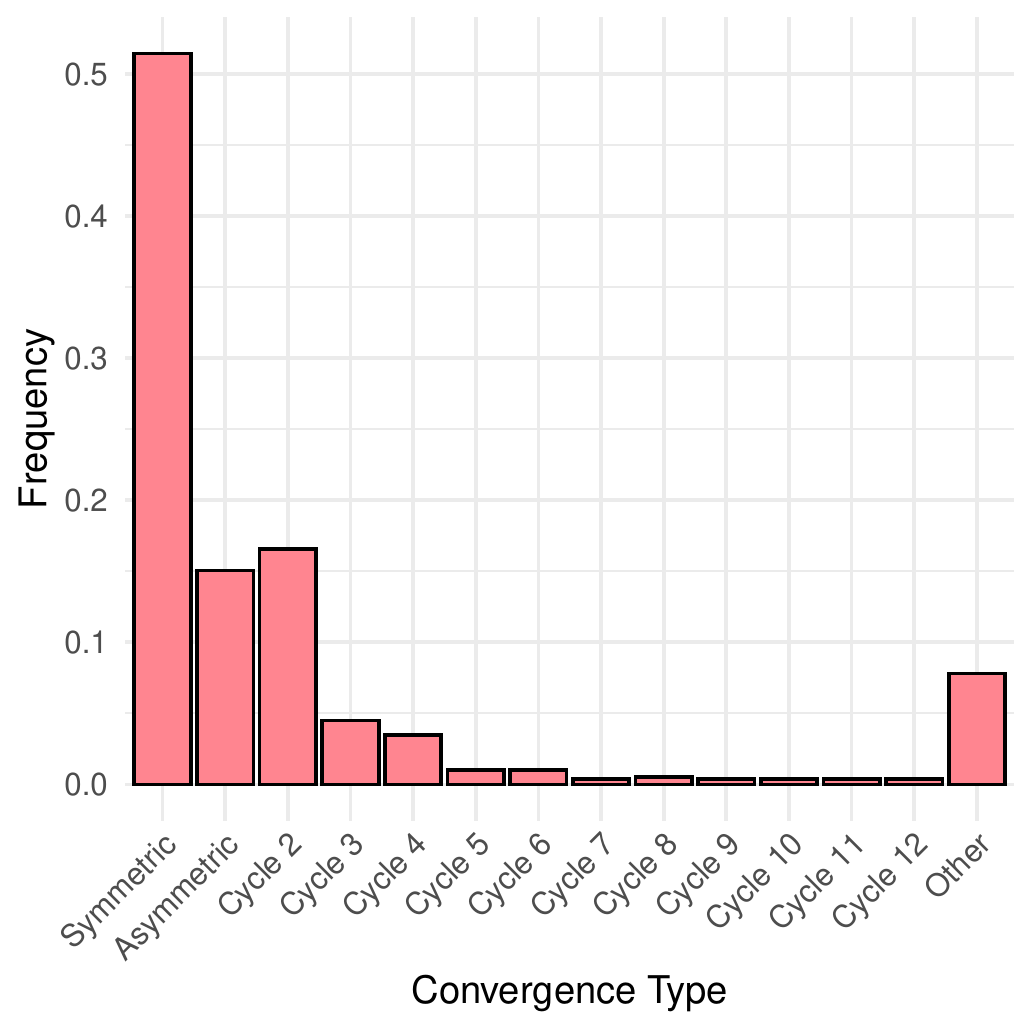}
  \end{subfigure}
\vspace{2em}
  \begin{subfigure}{0.495\textwidth}
    \caption{Off-diagonal testing context}
    \includegraphics[width=\linewidth]{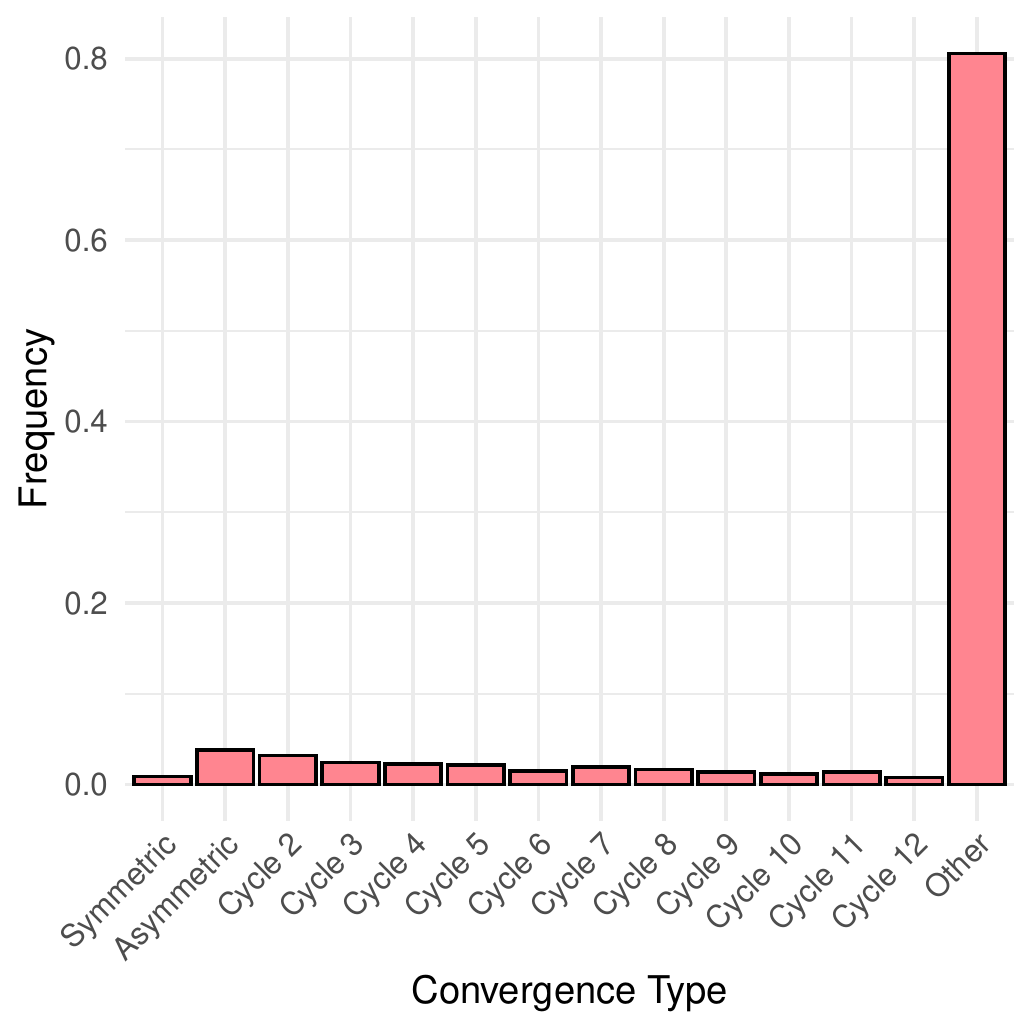}
  \end{subfigure}
\vspace{2em}
  \begin{subfigure}{0.495\textwidth}
    \caption{On-diagonal after re-convergence}
    \includegraphics[width=\linewidth]{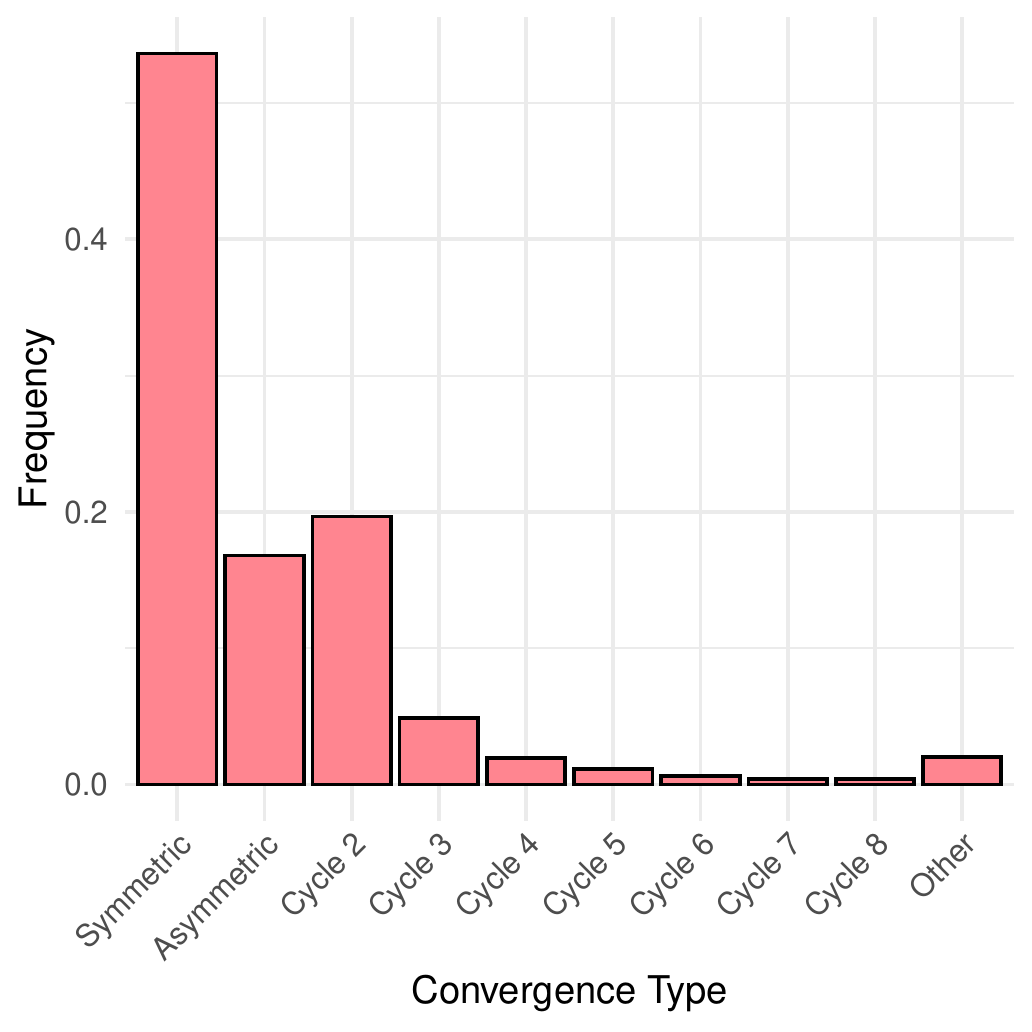}
  \end{subfigure}
  \begin{subfigure}{0.495\textwidth}
    \caption{Off-diagonal after re-convergence}
    \includegraphics[width=\linewidth]{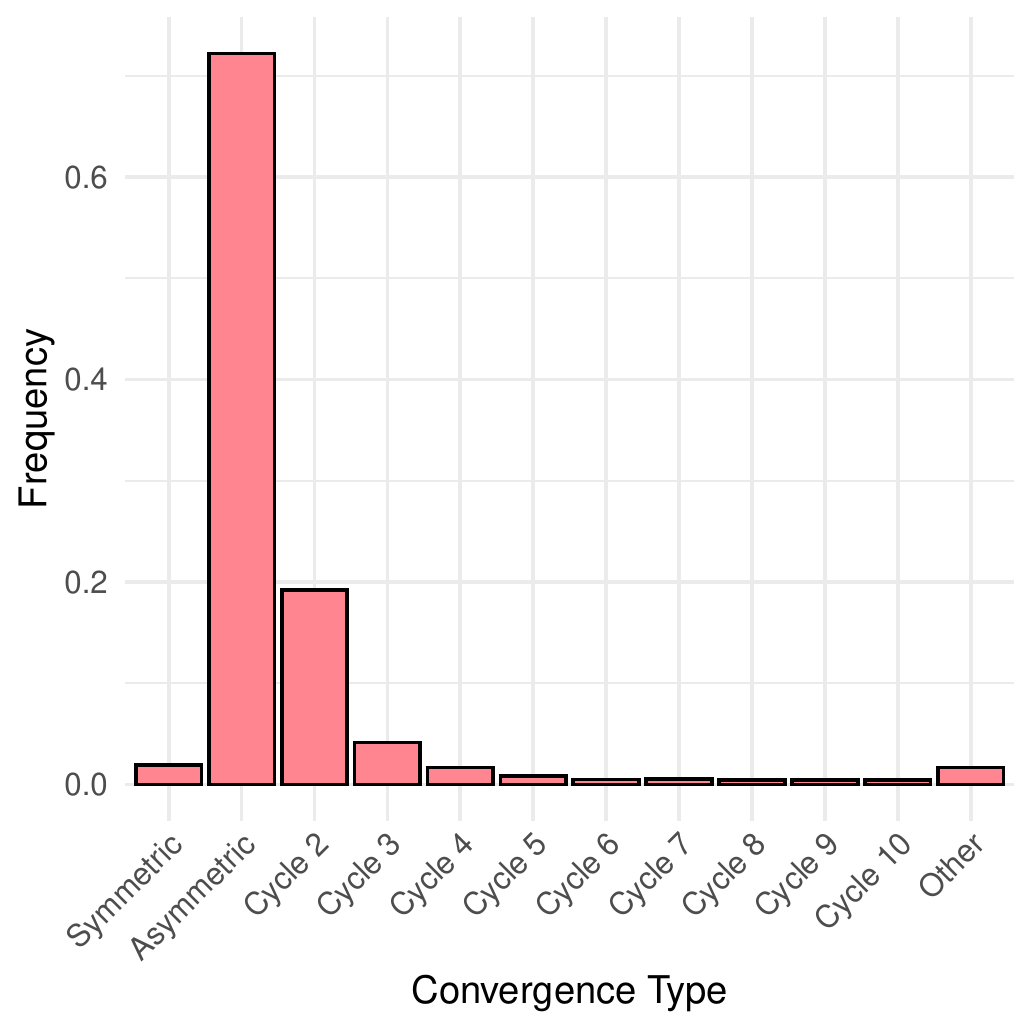}
  \end{subfigure}\\ \vspace{-1.5em}
  \flushleft{\footnotesize\textbf{Notes:} The figure shows a histogram of the frequency of convergence types in training contexts (left column) and testing contexts with asymmetric parameterization (right column) in the first $1000$ iterations and after re-convergence.}
\end{figure}

\begin{figure}[h!]
  \centering
  \caption{Example strategy pair with one stable point and collusive outcome}
  \includegraphics[width=0.9\columnwidth]{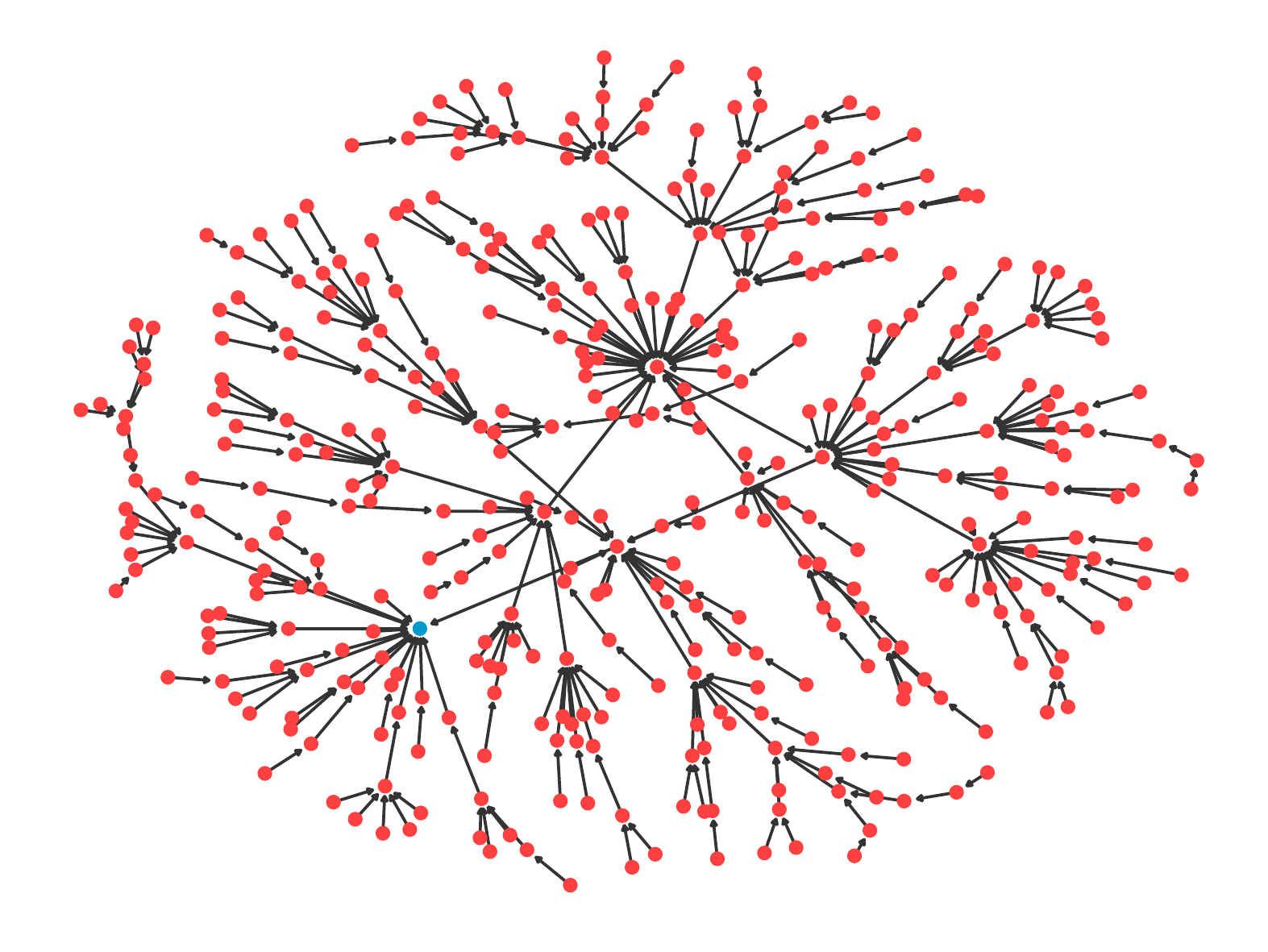}\\ \vspace{-1.5em}
  \flushleft{\footnotesize\textbf{Notes:} The figure shows the transitive closure of one pair of policies evaluated in the training context. Blue nodes are stable end-nodes, green nodes are unstable end-nodes.}
  \label{fig:network2}
\end{figure}

\begin{figure}[h]
\centering
\caption{Convergence types with restricted observation space and no updating}
\begin{subfigure}{0.49\columnwidth}
\subcaption{Training}
\includegraphics[width = \columnwidth]{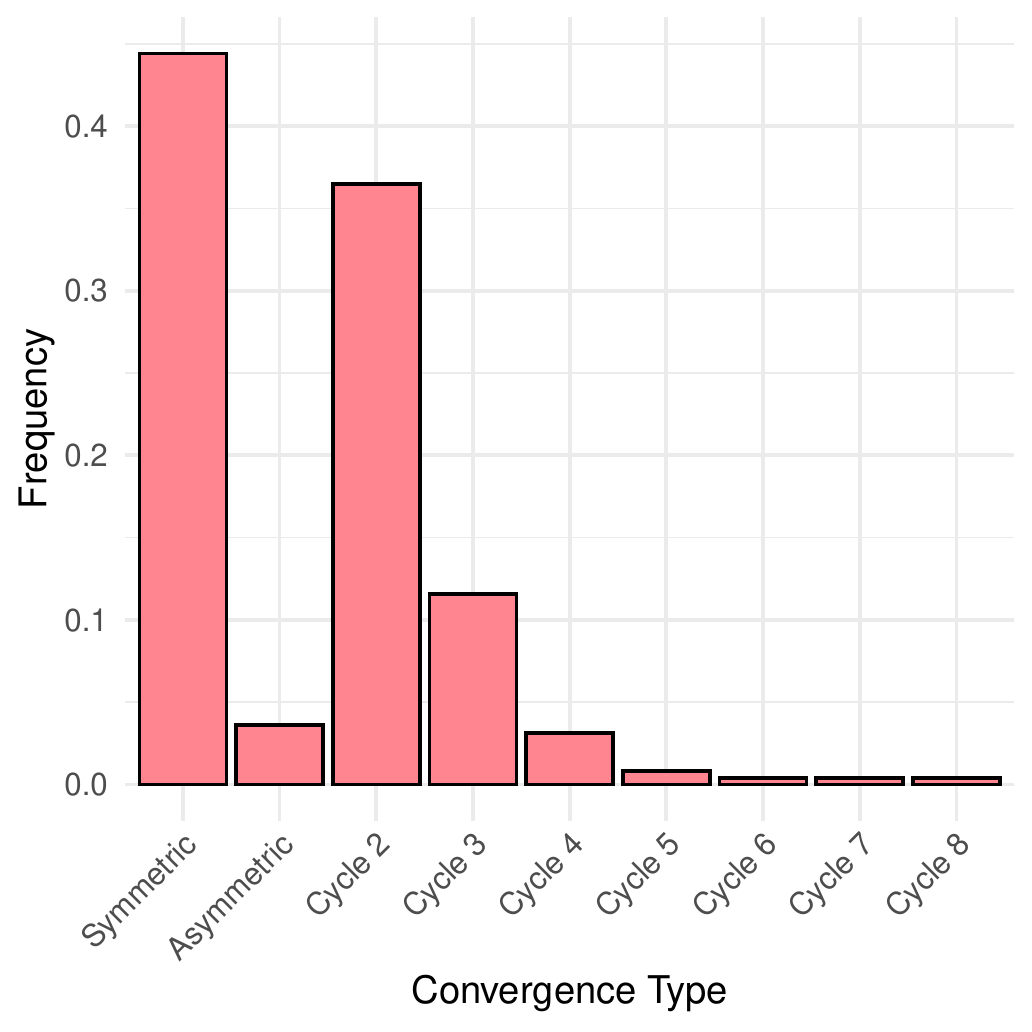}
\end{subfigure}
\begin{subfigure}{0.49\columnwidth}
\subcaption{Re-convergence}
\includegraphics[width = \columnwidth]{img/asymmetricLearnersOwnPriceObs-7b95b79f40f115bf3acff55918671d9e1191b512/equilibria_frequency_phase1.pdf}
\end{subfigure}\\ \vspace{-0.5em}
\flushleft{\footnotesize\textbf{Notes:} The figure shows a histogram of the frequency of convergence types in training contexts (panel (a)) and testing contexts after re-convergence (pabel (b)) in the setting with a restricted observation space.}
\label{fig:convergenceownpricetypes}
\end{figure}

\begin{figure}[h]
\centering
\caption{Convergence types with restricted observation space and updating}
\begin{subfigure}{0.49\columnwidth}
\subcaption{Training}
\includegraphics[width = \columnwidth]{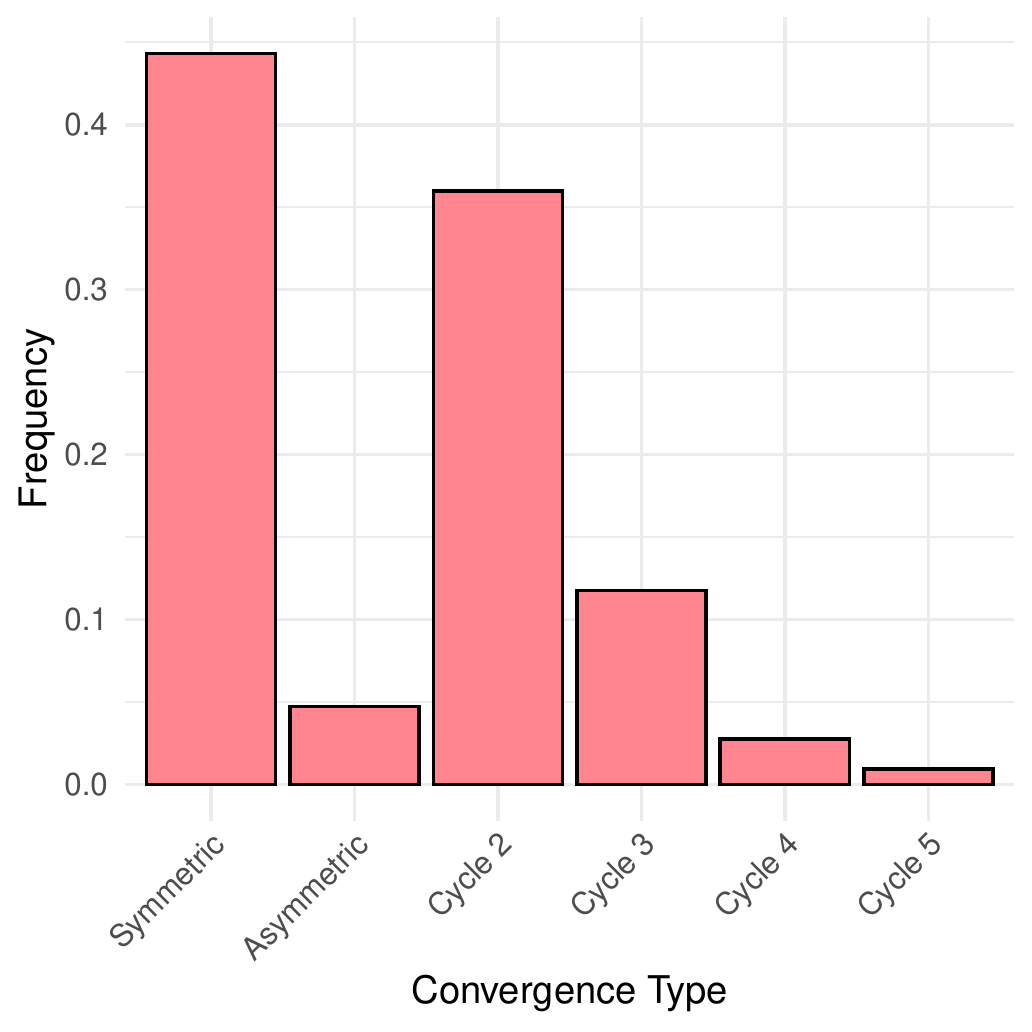}
\end{subfigure}
\begin{subfigure}{0.49\columnwidth}
\subcaption{Re-convergence}
\includegraphics[width = \columnwidth]{img/asymmetricLearnersOwnPriceObsLearning-6ae369612d59bf4d5b204e64cb880655ca1fd30d/equilibria_frequency_phase1.pdf}
\end{subfigure}\\ \vspace{-0.5em}
\flushleft{\footnotesize\textbf{Notes:} The figure shows a histogram of the frequency of convergence types in training contexts (panel (a)) and testing contexts after re-convergence (pabel (b)) in the setting with a restricted observation space when policies continue to be updated.}
\label{fig:convergenceownpriceupdatingtypes}
\end{figure}

\begin{figure}[h!]
  \centering
  \caption{Average path of play following deviation to Nash with restricted observation space}
  \includegraphics[width=0.9\columnwidth]{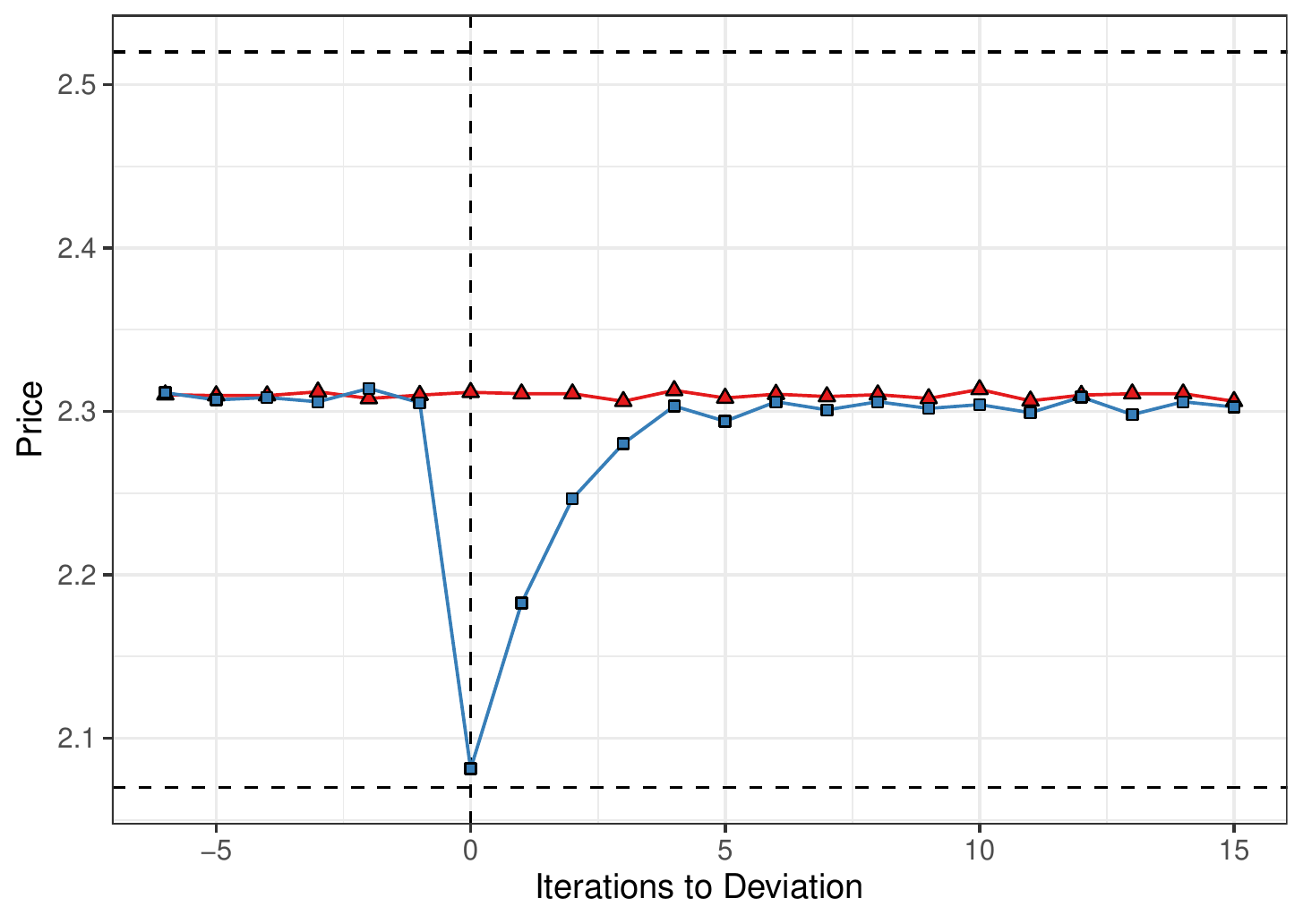}\\ \vspace{-1.5em}
  \flushleft{\footnotesize\textbf{Notes:}The figure shows the average actions played prior to and following a manual deviation of one player ({\color{blue}{in blue}}) to the price closest to the Nash price on the grid for training contexts with high cost parameterization with a restricted observation space.}
  \label{fig:avgdeviationownprice}
\end{figure}

\clearpage

\section{Tables}
\setcounter{table}{0}
\renewcommand{\thetable}{\Alph{section}\arabic{table}}

\begin{table}[h]
\begin{threeparttable}
\caption{Summary Statistics Training}
\label{tab:sumstatsphase1}
\centering
\renewcommand{\arraystretch}{1.3}
\begin{tabular}{@{}l@{\hspace{0.5cm}}cccccccc@{}}
  \toprule
  & & &  \multicolumn{6}{c}{Convergence Type} \\
  \cline{4-9}
  & \multicolumn{2}{c}{Collusion Index} & & & \multicolumn{4}{c}{Cycle}\\
  \cline{2-3} \cline{6-9}
  Cost Level & Mean  & SD & Symmetric & Asymmetric & 2 & 3 & 4 & 5+ \\
  \midrule
  1.00 & 0.79 & 0.16 &  79 &  52 &  59 &  27 &  23 &  10 \\
  1.10 & 0.81 & 0.13 &  89 &  51 &  65 &  26 &  12 &   7 \\
  1.20 & 0.87 & 0.11 & 159 &  20 &  39 &  15 &   9 &   8 \\
  1.30 & 0.87 & 0.10 & 171 &  23 &  36 &   9 &   4 &   7 \\
  1.40 & 0.84 & 0.10 & 168 &  36 &  42 &   3 &  0  &   1 \\
  1.50 & 0.75 & 0.12 & 164 &  49 &  32 &   5 &  0  &   0 \\
  1.60 & 0.70 & 0.13 & 156 &  50 &  38 &   5 &   1 &   0 \\
  1.70 & 0.73 & 0.15 & 154 &  52 &  40 &   3 &   1 &   0 \\
  \bottomrule
\end{tabular}
\begin{tablenotes}[flushleft]
   \item \textbf{Notes:} The table shows the mean and standard deviation of the collusion index averaged over training contexts per parameterization, as well as the frequency of convergence types in training contexts.
  \end{tablenotes}
\end{threeparttable}
\end{table}

\begin{table}[h]
\begin{threeparttable}
\caption{Profit gain or loss in percentage of Nash-equilibrium profit}
\label{tab:costoflearning}
\centering
\renewcommand{\arraystretch}{1.3}
\begin{tabular}{@{}l@{\hspace{1cm}}c@{\hspace{1cm}}c@{\hspace{1cm}}c@{}}
  \toprule
  &  & \multicolumn{2}{c}{Opponent Randomizes}\\
  \cline{3-4}
  Cost Level & Both Randomize &  Nash & Best-response \\
  \midrule
  1.00 & 0.13\% & 0.63\% & 0.80\% \\
  1.10 & 0.09\% & 0.50\% & 0.67\% \\
  1.20 & 0.02\% & 0.36\% & 0.53\% \\
  1.30 & -0.08\% & 0.32\% & 0.39\% \\
  1.40 & -0.21\% & 0.17\% & 0.25\% \\
  1.50 & -0.36\% & 0.04\% & 0.11\% \\
  1.60 & -0.54\% & -0.04\% & -0.02\% \\
  1.70 & -0.73\% & -0.17\% & -0.15\% \\
   \bottomrule
\end{tabular}
\begin{tablenotes}[flushleft]
   \item \textbf{Notes:} The table shows the difference in profit relative to the static Nash equilibrium profit per parameterization for thre cases.
  \end{tablenotes}
\end{threeparttable}
\end{table}

%

\clearpage


\end{document}